# Rovibrational dynamics in strongly-coupled gas-phase methane

Jane C. Nelson,[1] Trevor H. Wright,[1] Madeline Rohde,[1] Neo Lin,[1] and Marissa L. Weichman[1,*]

[1]Department of Chemistry, Princeton University, Princeton, NJ 08544, USA

[*]To whom correspondence should be addressed. E-mail: weichman@princeton.edu

## ABSTRACT

Polaritons, the hybrid light-matter states formed when molecular transitions are strongly coupled to an optical cavity mode, appear to alter chemical dynamics, reactivity, and photophysics, though the mechanisms and reproducibility of these effects remain hotly debated. Nearly all reports in polariton chemistry rest on spectroscopic measurements of optical cavity transmission. Interpretation of these experiments therefore relies on understanding the linear and nonlinear spectroscopy of strongly-coupled systems. Working in the gas phase can dramatically simplify spectroscopic observables, as the intermolecular and environmental interactions that complicate the response of condensed-phase systems can be minimized. Our prior work introduced a platform for gas-phase molecular polaritons and demonstrated that linear cavity transmission spectra in the strong coupling regime are well described with classical Fabry-Pérot cavity physics. Here, we extend this platform to the nonlinear spectroscopy of cavity-coupled rovibrational transitions of methane. We monitor transient changes in the cavity transmission spectrum following optical pumping of intracavity molecules via an axis orthogonal to the longitudinal cavity axis. We find that the classical Fabry-Pérot cavity treatment quantitatively reproduces our transient spectra and allows for extraction of the collisional relaxation and transport timescales of intracavity molecules. We observe no difference in the dynamics of strongly-coupled methane versus free-space, non-cavity control experiments. This effort connects our gas-phase work to the much larger body of transient condensed-phase polariton spectroscopy and represents a step towards predictive understanding of molecular dynamics under strong coupling.

## INTRODUCTION

Physical chemists have pursued the goal of directing chemical transformations with light since the invention of the laser.[1,2] Impressive efforts in laser-driven chemistry – with, for instance, vibrational mode-specific,[3] stereodynamic,[4] and coherent control[5] techniques – have proved this ideal possible in limited small-molecule systems. Over the past decade, the strong coupling of molecular transitions to the confined electromagnetic field of an optical cavity has emerged as a potential new approach for photonic control of chemical phenomena.[6–8] The strong light-matter coupling regime is characterized by the formation of hybrid light-matter states dubbed polaritons. Reports of polariton-altered molecular behavior have generated a massive theoretical effort,[9–11] many review articles,[6–8,12–14] and ongoing debate over mechanisms, interpretation, and reproducibility.[15–24] The vast majority of experimental polariton chemistry experiments use optical spectroscopy, most often cavity transmission spectroscopy, to track intracavity molecular processes. Progress in the field therefore hinges upon reaching a rigorous understanding of the linear and nonlinear spectroscopic observables encountered in these systems. Here, we report on nonlinear pump-probe spectroscopy of strongly-coupled gas-phase methane, building on our previously reported experimental platform for linear spectroscopy in the same system.[25–27] We demonstrate that transient cavity transmission spectroscopy, like linear cavity transmission, can be described with quantitative accuracy using classical Fabry-Pérot cavity optics.

Polariton formation is conventionally introduced within the quantum optics framework. As described by the Jaynes-Cummings model, a quantized photonic cavity mode and an optically bright transition of a single molecule hybridize under sufficiently strong coupling to form upper and lower polariton states.[11,28,29] By definition, the strong coupling regime is achieved when the Rabi splitting ($\Omega_R$) between the polariton states exceeds both the bare molecular ($2\gamma$) and cavity ($2\kappa$) full-width-at-half-maximum (fwhm) linewidths (Fig. 1a). This splitting is typically observed spectroscopically in the cavity transmission spectrum. In the collective coupling regime described by the Tavis-Cummings model, the excitations of $N$ intracavity molecules hybridize with the cavity mode to yield a collective Rabi splitting that scales as $\sqrt{N}$.[30] A manifold of $N-1$ dark states is also formed from non-totally symmetric linear combinations of the $N$ molecular excitations; how these states differ from uncoupled molecules remains the subject of some discussion.[18,31]

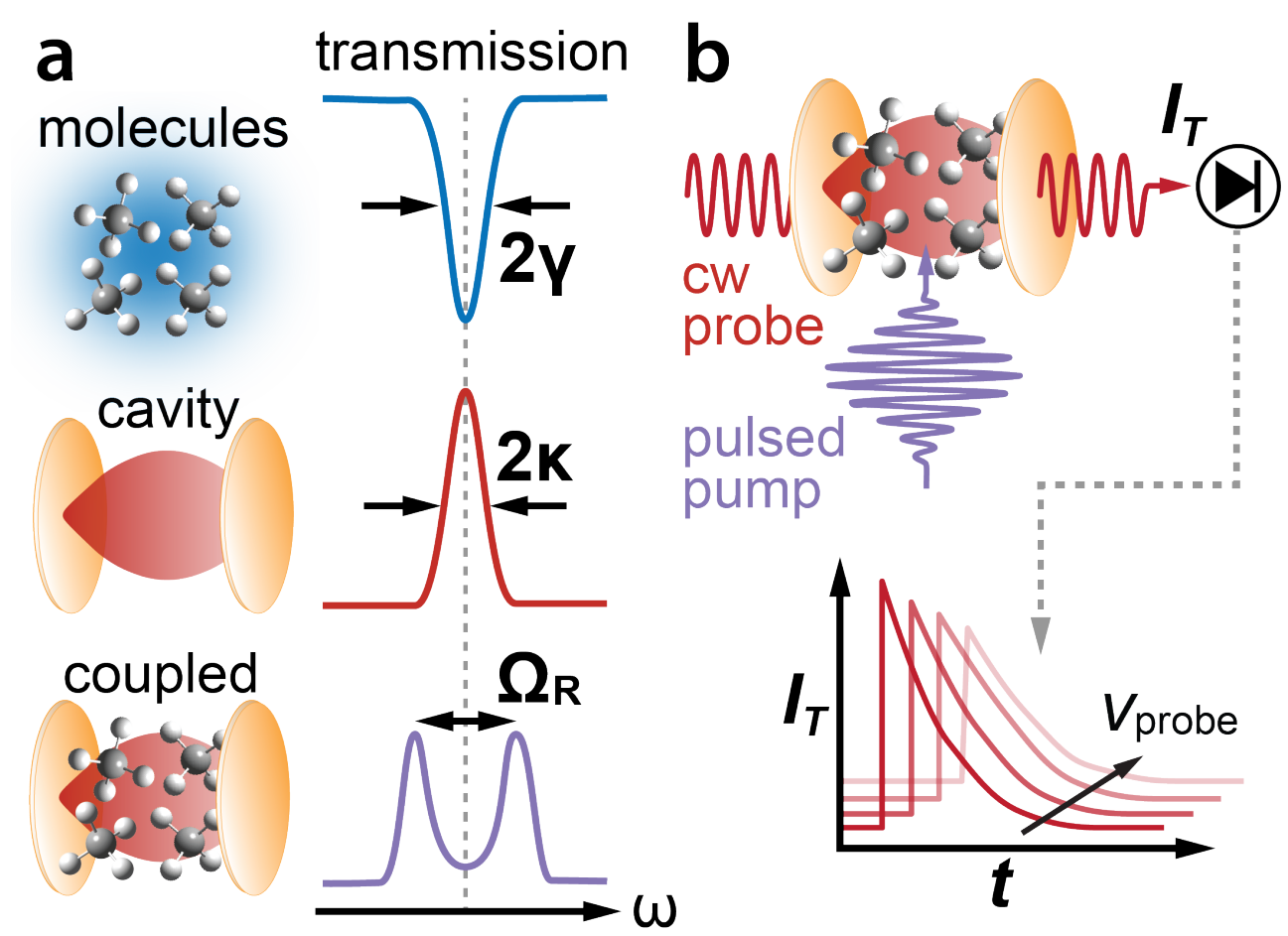

**Fig. 1. (a)** Spectroscopic signatures of a polaritonic system. Under strong light-matter coupling, the characteristic Rabi splitting (purple, $\Omega_R$) in the cavity transmission spectrum exceeds the fwhm linewidths of the original molecular (blue) and cavity (red) modes. **(b)** For nonlinear experiments, a nanosecond pump pulse aligned orthogonally to the cavity axis drives intracavity molecular population out of the lower state of the strongly-coupled transition. A continuous-wave (cw) probe laser records the intensity of light transmitted through the cavity, $I_T$, as a function of time following the pump modulation, at systematically stepped probe laser frequencies ($\nu_{\text{probe}}$).

While the quantum optics treatment underlies most theoretical work in the field and motivates the possibility of emergent cavity-altered chemical phenomena, it is not necessary to invoke quantum light in order to understand polaritonic cavity transmission spectra. In practice, molecular polariton spectra under collective strong coupling can be modeled by treating the hybrid light-matter system as a Fabry-Pérot (FP) cavity composed of two identical mirrors and filled with a homogeneous dielectric medium.[13,27,32–34] We have so far found this approach to provide excellent agreement with experiment for a wide array of strongly-coupled systems studied in our group.[25,35–38] Empty optical cavities feature transmission peaks evenly spaced by the free spectral range, $FSR = c / 2nL$, where $L$ is the cavity length, $n$ is the intracavity refractive index, and $c$ is the speed of light. The dispersion of a sufficiently strongly-absorbing intracavity medium causes a resonant cavity mode to split into two transmission peaks spaced by the same collective Rabi splitting predicted by quantum optics. Much ongoing debate in the field comes down to whether a given experimental report of intracavity molecular behavior is better interpreted with the quantum optics or classical optics treatment.

Strong light-matter coupling was first established in gas-phase atoms[39–41] building on early demonstrations of enhanced rates of spontaneous emission under weak coupling.[42] The emergent phenomena observed in atomic systems inspired subsequent exploration of exciton polaritons in solid-state semiconductor platforms.[33,43–45] Strong cavity-coupling has since been extended to the electronic and vibrational transitions of molecules in films and in solution,[6,13,46,47] inviting the possibility of using cavity coupling as a lever to control chemical reactivity and photophysics. Many recent experimental studies have focused on solution-phase reaction kinetics under vibrational strong coupling (VSC) in microfluidic cavities,[12,13] where early evidence of altered reactivity[47–50] has been followed by reports of non-reproducibility,[15,16] null results,[36,51,52] and optical artifacts.[20,24]

In parallel, transient pump-probe and two-dimensional infrared spectroscopy experiments have aimed to understand the dynamics and relaxation pathways of both polaritons and dark states in systems under VSC.[14,53–58] Much of this work has observed surprisingly long-lived signals following direct optical pumping of vibrational polaritons, leading to debate over whether these signals arise from the polaritons themselves or from the response of the intracavity molecular reservoir filtered by the cavity transmission spectrum.[21,53,56,59–62] In any event, it is clear that transient cavity transmission represents a profoundly distinct observable from transient absorption in free space. Transient cavity signals arise from both the intracavity molecular response *and* how the cavity's optical interference conditions change in reaction to that molecular response. The coupled molecular and cavity contributions can be disentangled with the help of the classical optics formalism,[21,37] though one must still take care that optical artifacts do not confound this approach.[20,24]

Quantitative simulation of nonlinear polariton spectroscopy is particularly challenging in condensed-phase systems. For one, the geometric constraints of micron-scale FP cavities necessitate that pump and probe pulses must be introduced through the cavity mirrors, leading to spectral filtering of light by the cavity transmission spectrum and temporal broadening according to the photonic cavity lifetime.[18,21,34] In addition, condensed-phase molecules experience significant interactions with other molecules and the solvent bath which can dominate their dynamics. Condensed-phase VSC experiments also typically involve cavity-coupling of entire vibrational bands with unresolved inhomogeneous structure and congestion from neighboring bands.[21] All of these effects complicate the simulation and analysis of the coupled cavity-molecule system's transient response.

By contrast, the gas phase offers several advantages for studying polariton dynamics. Intracavity molecular density, temperature, and background pressure can be easily tuned, enabling careful control over polaritonic properties. Intermolecular interactions are minimized, so the coupling of target molecules to the bath no longer dominates linear spectra or transient dynamics. Rovibrationally-resolved transitions are narrow and sparse, permitting clean, state-specific coupling even in cavities with densely spaced longitudinal modes. This last point allows us to use centimeter-scale curved-mirror FP cavities that feature smaller *FSR*s and narrower linewidths than micron-scale geometries. Orthogonal access to intracavity molecules becomes possible in these longer cavities, circumventing the complications of spectral filtering of on-axis pump pulses. Furthermore, mode matching of light into the Gaussian transverse electromagnetic (TEM) modes supported by curved-mirror cavities eradicates concerns over spatial inhomogeneity in cavity length that can greatly complicate measurements in planar microcavities.[20,24]

Here, we report the first time-resolved pump-probe measurements of the dynamics of gas-phase molecules under VSC, harnessing our previously reported platform for cavity coupling of the individual rovibrational transitions of methane.[25–27] We now introduce a nanosecond pulsed pump beam along an axis orthogonal to the longitudinal cavity axis to directly pump the intracavity molecular population by exciting the same rovibrational transition that is strongly coupled. The system's transient response is read out via the time-resolved transmission of a narrow-band continuous-wave (cw) probe laser aligned on-axis through the cavity (Fig. 1b). We simulate the experimental transient cavity transmission spectra with quantitative accuracy using an extension of the classical FP cavity optics framework widely used to describe the linear cavity transmission spectroscopy of strongly-coupled systems. These simulations allow us to recover the intracavity molecular response that gives rise to transient cavity transmission signatures, which can then be compared directly to non-cavity control experiments. We find all intracavity molecular dynamics to be consistent with non-cavity controls, with timescales following those expected for the collisional relaxation and transit times of methane. We establish that, for this system, classical FP cavity optics provides as quantitative a treatment of nonlinear spectroscopy under strong coupling as is well accepted for linear spectroscopy.

## EXPERIMENTAL METHODS

Our apparatus for linear spectroscopy of gas-phase molecular rovibrational polaritons has been described in detail in previous reports;[25,26] the setup used here (Fig. 2) is a modification of that existing apparatus. As in our prior work, we target strong coupling of the $J = 3 \rightarrow 4$ multiplet in the R branch of the $\nu_3$ C-H stretching fundamental of methane near 3270.445 nm (3057.687 $cm^{-1}$). We hereafter refer to this transition as $\nu_3$ R(3).

Following our established protocol,[25,26] we prepare polaritons by flowing methane gas into a centimeter-scale FP optical cavity. We lock the cavity length to keep a single cavity mode resonant with the selected rovibrational transition of methane. We record the cavity transmission spectrum using a narrowband cw laser and observe the characteristic Rabi splitting when the intracavity number density of methane is sufficiently high. In the present work, we introduce a nanosecond pulsed pump laser orthogonal to the longitudinal cavity axis that can be tuned to resonantly excite the same rovibrational transition of methane that is strongly coupled. The intensity of cw probe laser light transmitted through the longitudinal cavity axis is then recorded as a function of time following the pump pulse. By systematically stepping the cw probe laser, we construct the transient cavity transmission signals as a function of both frequency and delay time. Additional experimental details are provided in Section S1 of the Supporting Information (SI).

Strong coupling experiments are performed in a vacuum chamber into which we flow methane gas (Fig. 2a). We evacuate the chamber with a turbomolecular vacuum pump backed by a scroll pump. Methane gas is delivered to the center of the chamber through a copper gas line regulated by a mass flow controller; no additional buffer gas is added in the current experiments. In our prior work,[25,26] we cooled the methane gas in an intracavity cryogenic buffer gas cell mounted to a liquid nitrogen ($LN_2$) dewar. For the present experiments, we remove the buffer gas cell to allow a shorter cavity with accordingly longer *FSR* to reduce spectral congestion. We work at room temperature for most of the results reported herein, though we can still fill the dewar with $LN_2$ to yield some measure of gas cooling. The copper gas delivery line is wound around a screw secured to the $LN_2$ dewar to pre-cool the gas for cold experiments. The temperature and pressure conditions for the cooled chamber are detailed in Section S1.1 of the SI.

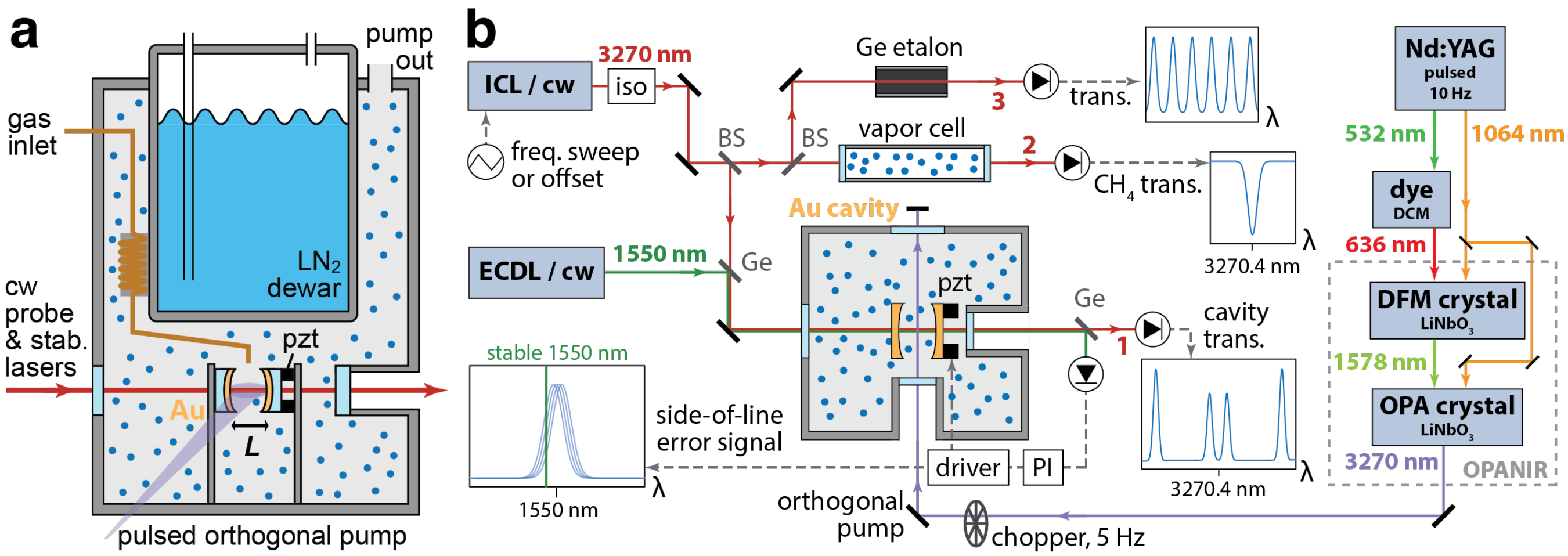

**Fig. 2. (a)** Vacuum chamber housing the Fabry-Pérot cavity used to strongly-couple gas-phase molecules. Gas is delivered to the cavity via a copper line; for cold experiments, the gas is pre-cooled as the copper line is in thermal contact with a liquid nitrogen ($LN_2$) dewar. Plano-concave Au cavity mirrors are mounted to the base of the chamber and separated by length $L$. Continuous-wave (cw) lasers are aligned along the longitudinal cavity axis for transmission spectroscopy and cavity length stabilization. The pulsed pump laser is aligned orthogonal to the cavity. **(b)** Laser scheme used to record cavity transmission spectra, perform absolute and relative frequency calibration, stabilize the optical cavity length, and pump the intracavity sample. 3270 nm light from an interband cascade laser (ICL, red) passes through an optical isolator (iso) and is then split into three arms with beamsplitters (BS). The ICL is mode-matched into the cavity $TEM_{00}$ mode for cavity transmission spectroscopy (beam 1). A $CH_4$ vapor cell serves as an absolute frequency reference (beam 2), while a solid germanium (Ge) etalon is used for relative frequency calibration (beam 3). A 1550 nm external cavity diode laser (ECDL, green) is used to stabilize the cavity length via a side-of-line lock. The 1550 nm light is coupled in and out of the cavity using Ge flats as longpass dichroic mirrors. The side-of-line error signal is processed with a proportional-integral (PI) loop filter and fed back onto the cavity length via a piezoelectric chip (pzt). To conduct time-resolved measurements, pulsed 3270 nm pump light is generated at 10 Hz using a commercial optical parametric amplifier near-infrared (OPANIR) laser system with difference frequency mixing (DFM) and OPA stages. The OPANIR is pumped with a dye laser and an Nd:YAG laser.

We probe the cavity transmission with a mid-infrared distributed feedback interband cascade laser (ICL) with an instantaneous linewidth of a few MHz and a wavelength centered near 3270.4 nm where the target transitions of $CH_4$ lie (Fig. 2b). The ICL can be tuned over several wavenumbers via a voltage applied to the modulation port of its current driver. We operate the ICL in either a frequency-swept mode with a triangle wave modulation to acquire linear cavity transmission spectra or a frequency-stepped mode with a DC voltage offset applied to acquire the time-resolved cavity transmission signal at a fixed frequency. The majority of the ICL light is mode-matched into the $TEM_{00}$ mode of the cavity with a pair of plano-convex lenses (beam 1 in Fig. 2b). This beam has a vertical linear polarization set with a half-wave plate. Optical access to the chamber is made through anti-reflection coated $CaF_2$ windows; these windows are recessed where possible to minimize absorption by extracavity gas. We record the intensity of light transmitted through the cavity with an $LN_2$-cooled HgCdTe detector with a nominal 53 ns response time. In parallel, a small portion of the ICL light is picked off to pass through a reference $CH_4$ gas cell to an InAsSb detector for absolute frequency calibration (beam 2 in Fig. 2b) and through a solid Ge etalon to a second InAsSb detector for relative frequency calibration (beam 3 in Fig. 2b). Details on the frequency calibration process are provided in Sections S1.2 and S2.1 of the SI.

The FP cavity used to achieve strong coupling is configured in a near-confocal geometry with cavity length $L$ set to match the mirror radii of curvature ($ROC$) of –1.05 cm. Confocal cavities feature highly-degenerate even and odd transverse spatial modes spaced by $c/4nL$, half of the typical FP cavity $FSR$.[63] Here, we intentionally mode match our probe laser to the $TEM_{00}$ spatial mode, yielding a cavity transmission spectrum with fringes spaced by the usual $FSR = c/2nL \sim$ 14.3 GHz expected for an $L \sim 1.05$ cm cavity. The cavity beam waist ($1/e$ radius) for the $TEM_{00}$ mode of this cavity is ~74 µm for light with wavelength near 3270 nm.[63] The cavity mirrors consist of 20 nm of gold deposited with an electron beam evaporator on ½” diameter plano-concave $CaF_2$ substrates. The mirrors are mounted in standard optics mounts clamped to the base of the chamber. One mirror is glued to a ring-shaped piezoelectric chip (pzt) to enable remote actuation of the cavity length.

In all experiments performed herein, the cavity length is actively stabilized using a side-of-line lock to a metrology-grade cw external cavity diode laser (ECDL) with a central wavelength near 1550 nm. Briefly, the ECDL transmission intensity through a cavity mode is processed as an error signal with a proportional-integrator loop filter and fed onto the cavity pzt to actuate the cavity length. The circuit works to counteract thermal drifts in cavity length and fix the frequencies of the cavity modes both near the 1550 nm lock point and near 3270.4 nm where we investigate strong coupling. This locking scheme has been described in our prior work;[25,26,35] more details are also provided in Section S1.3 of the SI.

We introduce a nanosecond pulsed pump laser to modulate the population of intracavity molecules in the lower level of the transition targeted for strong coupling. This laser system is illustrated in Fig. 2b and explained in more detail in SI Section S1.4. In short, we generate ~6 ns pulses of mid-IR light at 10 Hz using a commercial optical parametric amplifier near-infrared (OPANIR) frequency conversion module fitted with two $LiNbO_3$ crystals (Sirah Lasertechnik). The OPANIR is specified to produce mid-IR light wavelength-tunable from 2900−4000 nm with a <0.066 $cm^{-1}$ (2.0 GHz) fwhm linewidth. To produce pump light near the targeted $\nu_3$ R(3) transition of methane at 3270.4 nm, the OPANIR is pumped with 636 nm light from a dye laser circulating DCM dye in ethanol (Sirah Lasertechnik) and the 1064 nm fundamental of an Nd:YAG laser (Amplitude). For a subset of experiments, we set the dye laser to 633 nm to instead pump the $\nu_3$ P(3) transition of methane at 3345.8 nm. The dye laser is itself pumped with the 532 nm second harmonic of the same Nd:YAG. We determine the mid-IR wavelength by measuring the Nd:YAG fundamental and dye laser wavelengths to within <0.001 nm using a wavemeter (Bristol Instruments).

The pulsed mid-IR pump beam is collimated after the OPANIR and aligned through the vacuum chamber orthogonal to the cavity axis via a recessed, anti-reflection coated $CaF_2$ window. Approximately 3 mJ/pulse of pump light reaches the chamber. We focus the pump light using a cylindrical lens to achieve a 1 cm × 200 μm beam profile that permits uniform excitation over the cavity mode volume. A mechanical chopper blocks every other pump pulse to enable pump-on/pump-off background correction from neighboring pairs of laser pulses (*vide infra*). Like the cavity-transmission probe beam, the pump beam is vertically polarized.

Before each time-resolved experiment, we collect a reference linear cavity transmission spectrum $I_{ref}(\nu)$ by sweeping the frequency of the ICL probe laser over a 30 GHz window of interest to characterize the initial cavity and molecular conditions. We calibrate the frequency axis of $I_{ref}(\nu)$ using the etalon and vapor cell reference spectra acquired synchronously, as described in Section S1.2 of the SI. We fit $I_{ref}(\nu)$ to extract cavity parameters and initial methane conditions as detailed in Section S2.1 of the SI. We then collect transient cavity transmission signals as a function of both frequency and time following pump-probe overlap at $t = 0$. We systematically step the ICL probe frequency in 33 MHz steps over a 1.7−2.5 GHz window centered around the $\nu_3$ R(3) resonance. At each ICL frequency setpoint we collect a series of 20 or 40 μs-long cavity transmission time traces, with 2 or 4 ns time sampling respectively, beginning acquisition ~1 μs before the arrival of the pump pulse. We collect 50−75 pairs of pump-on/pump off traces at each ICL frequency setpoint, discarding traces when the ICL probe laser is unstable or saturated; see Section S2.2 of the SI for more details. The time traces collected at a given ICL frequency setpoint $\nu_n$ are averaged together and smoothed to generate time-dependent cavity transmission traces with pump on [$I_{\text{on}}(\nu_n, t)$] and pump off [$I_{\text{off}}(\nu_n, t)$]. To determine the absolute ICL frequency at each setpoint, we map the baseline cavity transmission signal for $t < 0$ before the pump arrives onto the $I_{\text{ref}}(\nu)$ reference spectrum acquired for the same $[CH_4]_{\text{off}}$ condition. We define the pump-probe overlap time $t = 0$ when $d[I_{\text{on}}(\nu_n, t)]/dt$ exceeds a predetermined noise floor, as described in more detail in Section S2.2 of the SI.

We finally calculate the time-dependent differential cavity transmission at each ICL frequency setpoint $\nu_n$ according to:

$$\Delta\text{OD}(\nu_n, t) = -\log_{10}[\, I_{\text{on}}(\nu_n, t) \,/\, I_{\text{off}}(\nu_n, t) \,]. \tag{1}$$

We step $\nu_n$ to build up a full time- and frequency-resolved transient spectrum $\Delta\text{OD}(\nu, t)$. Because our data is so densely sampled in time, we ultimately bin $\Delta\text{OD}(\nu, t)$ into 10 or 20 ns time windows and average together the data within each bin. We hereafter label the $m^{\text{th}}$ of these discrete time bins as $t_m$.

## RESULTS AND DISCUSSION

We now examine the transient behavior of gas-phase methane under strong cavity coupling via nonlinear pump-probe spectroscopy. We record differential transient cavity transmission spectra in the region of a single strongly-coupled rovibrational transition, following optical excitation of population out of the lower state of the coupled transition. By fitting experimental results with classical FP cavity expressions, we extract the population dynamics of pumped intracavity molecules and directly compare their behavior to control experiments conducted in free space.

### Rovibrational strong coupling in $CH_4$

The mid-IR rovibrational transitions of methane are ideal targets for gas-phase cavity-coupling because of their strong absorption cross-sections, well-resolved line spacings, and narrow linewidths.[27] Here, we strongly-couple the brightest transition in the R(3) multiplet of methane's $\nu_3$ C−H stretching fundamental near 3270.4 nm (3057.7 cm$^{-1}$) by bringing a cavity mode into resonance with this transition. In most of the following transient experiments, we optically pump the same strongly-coupled R(3) transition, though we pump the $\nu_3$, $J$ = 3→2 P-branch transition at 3345.8 nm (2988.8 cm$^{-1}$), hereafter referred to as P(3), in a subset of experiments. Both R(3) and P(3) originate from the same $J$ = 3 rotational level of the ground vibrational state of methane (Fig. 3a).

We first characterize the uncoupled cavity and molecular transmission spectra. Figure 3b plots a representative empty-cavity transmission spectrum (red trace). This cavity features a $\Delta\nu$ = 278 MHz (0.009 cm$^{-1}$) fwhm photonic mode linewidth, which is, by design, similar in scale to the 280 MHz (0.009 cm$^{-1}$) fwhm room-temperature Doppler-broadened molecular linewidth as seen in the blue trace of Fig. 3b. Gaussian Doppler broadening dominates the molecular lineshape even at the highest flow conditions explored herein; with [$CH_4$] ~ 1.5 × 10$^{16}$ cm$^{-3}$, the self-broadened Lorentzian pressure broadening is still expected to be below 4 MHz fwhm.

We achieve strong cavity coupling by flowing methane into the chamber while keeping a cavity mode locked on resonance to the selected $\nu_3$ R(3) transition. Representative linear cavity transmission spectra, $I_{\mathrm{ref}}(\nu)$, are plotted in Fig. 4a for various intracavity methane concentrations. We observe a Rabi splitting of $\Omega_R$ = 554 MHz for a room temperature initial intracavity methane number density of [$CH_4$]$_{\mathrm{off}}$ = 8.0 × 10$^{15}$ cm$^{-3}$ (purple trace in Fig. 4a). The system enters the strong coupling regime for [$CH_4$]$_{\mathrm{off}}$ $\gtrsim$ 1.5 × 10$^{15}$ cm$^{-3}$ when the Rabi splitting exceeds the bare molecular and cavity linewidths near 280 MHz. We have demonstrated Rabi splittings as high as 888 MHz with [$CH_4$] = 4.2 × 10$^{16}$ cm$^{-3}$ in prior work.[26]

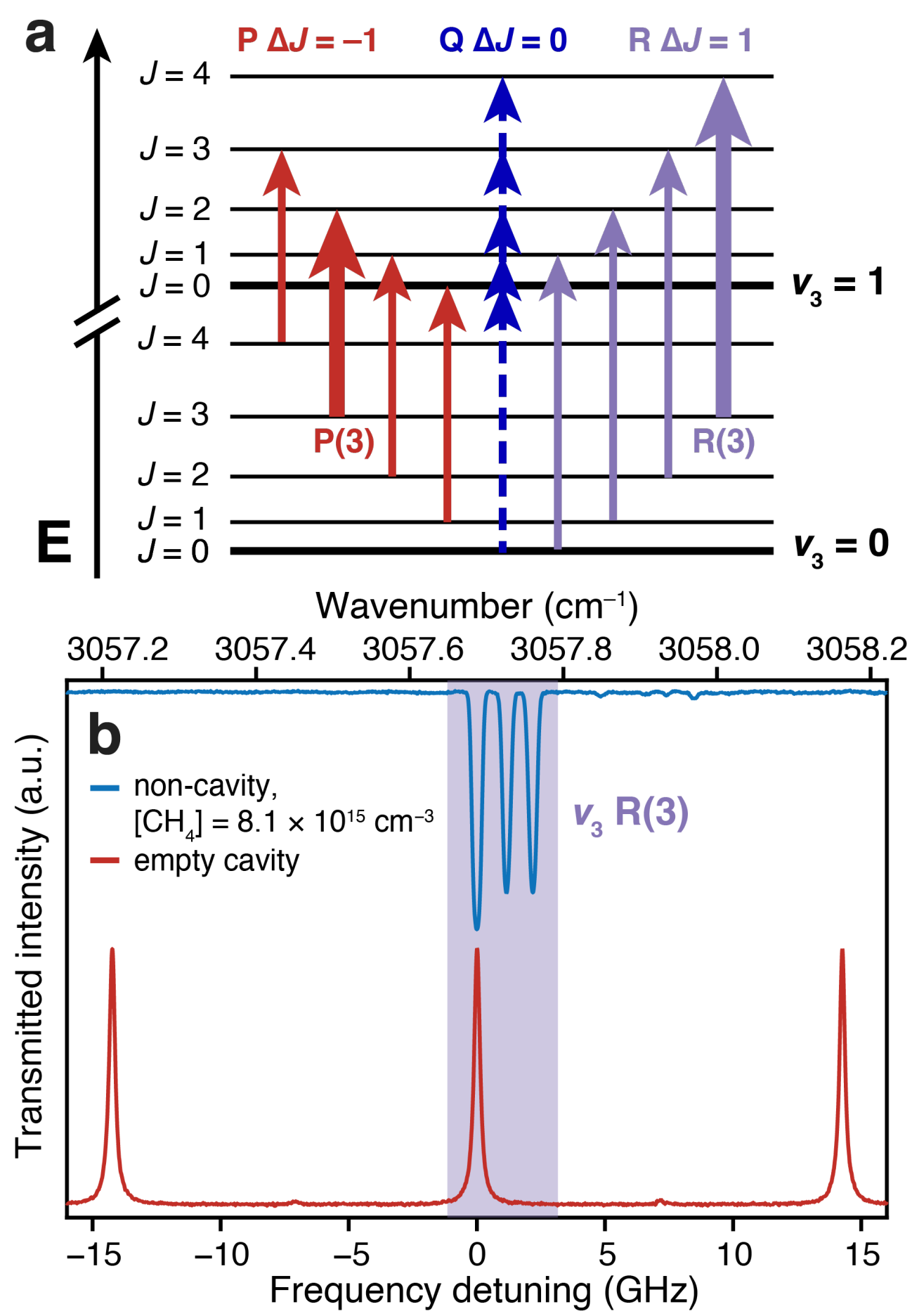

**Fig. 3. (a)** Energy level diagram of dipole-allowed P- (red, $\Delta J = -1$), Q- (blue, $\Delta J = 0$), and R-branch (purple, $\Delta J = +1$) rovibrational transitions for spherical top molecules. The $\nu_3$ R(3) transition (bold purple arrow) of methane is targeted for cavity coupling in this work. The P(3) transition (bold red arrow) originates from the same lower state as R(3), the $J = 3$ rotational level of the ground vibrational state. **(b)** Experimental transmission spectra of the near-confocal cavity used for strong coupling when empty (red) and of a free-space, non-cavity room temperature $CH_4$ sample (blue, vertically offset for clarity). The spectral region near the $\nu_3$ R(3) multiplet is shaded in purple. The frequency axis is referenced with the center of the primary $CH_4$ $\nu_3$ R(3) absorption band at 3057.687423 cm$^{-1}$ corresponding to 0 GHz.

### Modeling linear cavity transmission spectra

We model linear cavity transmission spectra using the analytical expression for the fraction of light transmitted through an FP cavity composed of two identical mirrors and filled with a homogeneous dielectric medium:[32,63]

$$\frac{I_T(\nu)}{I_0} = \frac{T^2 e^{-\alpha(\nu)L}}{1+R^2 e^{-2\alpha(\nu)L} - 2R\mathrm{e}^{-\alpha(\nu)L}\cos[\delta(\nu)]} \quad (2)$$

where $T$ and $R$ are the transmission and reflection intensity coefficients of the cavity mirrors, $L$ is the cavity length, $\alpha(\nu) = \sigma(\nu)\cdot[CH_4]$ is the absorption coefficient of the intracavity methane, $\sigma(\nu)$ is the absorption cross section, $n(\nu)$ is the refractive index, and $\delta(\nu) = 4\pi L n(\nu)\nu/c$ is the round-trip phase accrued by light traveling in the cavity, where $c$ is the speed of light. We simulate $\sigma(\nu)$ for the relevant rovibrational transitions in the $\nu_3$ band of methane using reference data available in the HITRAN database[64] processed in the software package PGOPHER[65] to capture experimentally-relevant Doppler and pressure broadening. We derive the refractive index $n(\nu)$ from $\alpha(\nu)$ via the Kramers-Kronig relation. We find that we must also use the Beer-Lambert law to account for the attenuation of ICL probe light as it traverses ~14 cm of extracavity pathlength in the chamber; see eq S1 and further discussion in Section S2.1 of the SI.

We fit each reference spectrum $I_{ref}(\nu)$ with eq 2, scaled for extracavity attenuation (eq S1), to extract $L$, $R$, and the initial molecular number density $[CH_4]_{off}$. The simulated fits of the experimental spectra shown in Fig. 4a are included in Fig. S1 in SI Section S2.1. Across all datasets reported herein, we find $L = 1.047 \pm 0.003$ cm and $R = 94.3 \pm 0.2\%$, resulting in a cavity finesse $\mathcal{F} \approx \pi\sqrt{R}/(1-R) = 53 \pm 2$ for $\lambda = 3270.4$ nm ($3057.7\ cm^{-1}$). Our experimental metered flow rates of 0 to 200 sccm of methane into the chamber within this work yield fitted number densities $[CH_4]_{off}$ ranging from 0 to $1.5 \times 10^{16}\ cm^{-3}$.

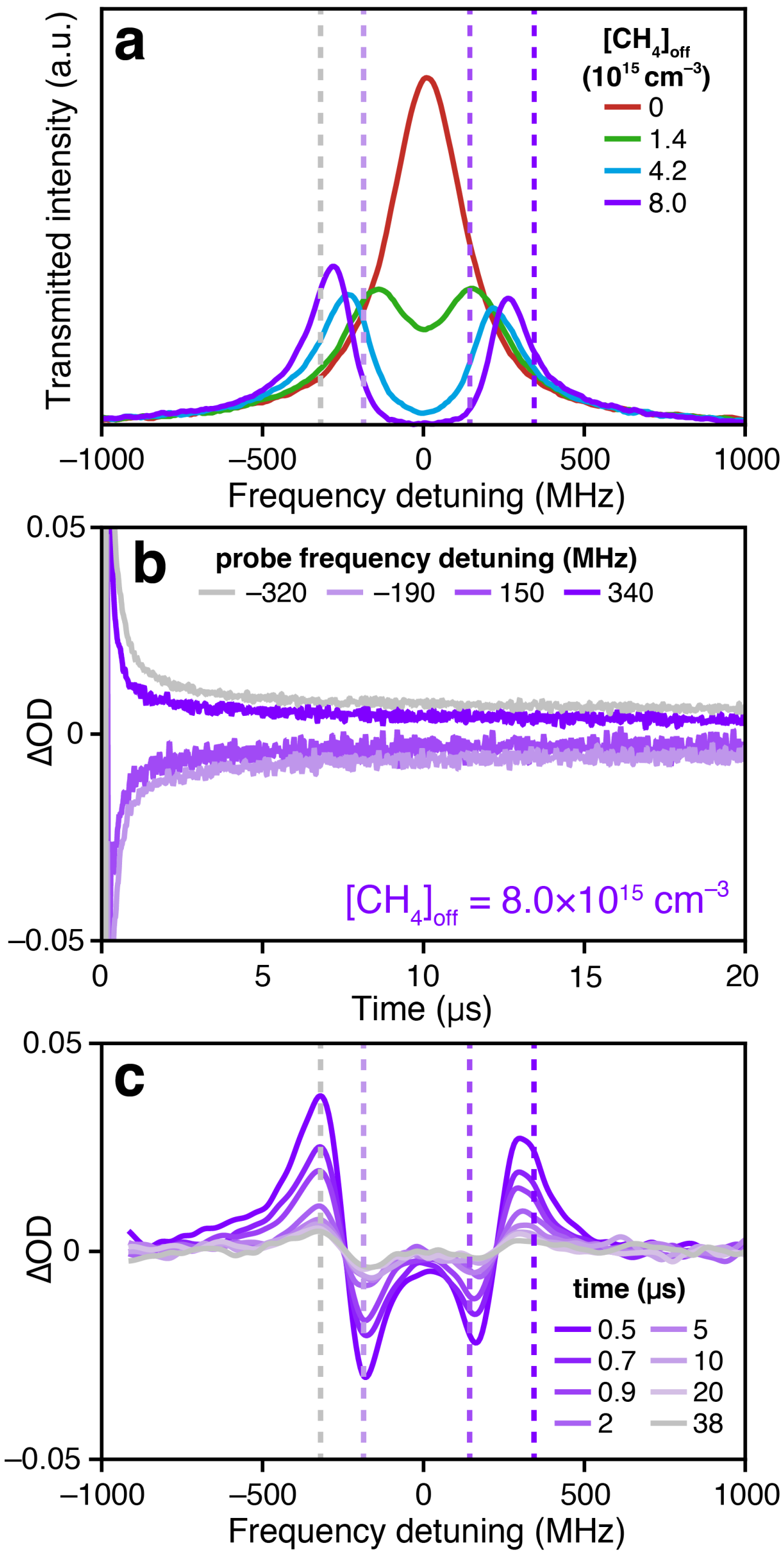

**Fig. 4. (a)** Experimental linear cavity transmission spectra, $I_{ref}(\nu)$, under systematic tuning of the initial intracavity methane number density $[CH_4]_{off}$ with flow rates of 0, 5, 25, and 75 sccm. $[CH_4]_{off}$ is determined from the cavity transmission spectral fitting routine described in Section S2.1 of the SI; fits are shown in Fig. S1. All data shown here are acquired at room temperature. **(b)** Differential cavity transmission time-traces at four representative ICL probe frequencies $\nu_n$ near the R(3) spectral region following optical pumping of R(3), calculated as $\Delta OD(\nu_n, t) = -\log_{10}[I_{on}(\nu_n, t)/I_{off}(\nu_n, t)]$. Data presented are for $[CH_4]_{off} = 8.0 \times 10^{15}$ cm$^{-3}$, corresponding to the purple trace from panel (a). **(c)** Differential cavity transmission spectra $\Delta OD(\nu, t_m)$ plotted as a function of frequency at various time points $t_m$. Color-coded dashed lines in panels (a) and (c) mark the frequency setpoints whose time traces are plotted in panel (b). The frequency detuning

axis in panels (a) and (c) is referenced with the center of the $CH_4$ $\nu_3$ R(3) absorption band at 3057.687423 $cm^{-1}$ corresponding to 0 MHz.

**Transient cavity transmission spectroscopy of strongly-coupled $CH_4$**

We perform nonlinear experiments by exciting the intracavity molecular sample with pump light resonant with the coupled $\nu_3$ R(3) transition and recording transient cavity transmission spectra $\Delta$OD($\nu$, $t$) as described above in the Experimental Methods section. Note that while we use the standard pump-probe "differential optical density" convention here, $\Delta$OD($\nu$, $t$) reflects pump-induced modulation of the cavity transmission rather than transient absorption. Representative transient time traces $\Delta$OD($\nu_n$, $t$) at various ICL frequency setpoints $\nu_n$ are plotted in Fig. 4b for $[CH_4]_{off} = 8.0 \times 10^{15}$ $cm^{-3}$. These frequency setpoints are highlighted with color-coded dashed lines in Figs. 4a,c. Frequency-domain differential spectra $\Delta$OD($\nu$, $t_m$) can then be constructed at various time points $t_m$, as plotted in Fig. 4c. We record transient changes in cavity transmission on the order of tens of mOD, a scale exceeding that typically achieved in condensed-phase ultrafast experiments. The differential lineshapes observed near the polariton mode frequencies are consistent with a Rabi contraction as molecules are pumped out of the $J$ = 3 rotational level of the vibrational ground state, the lower state of the strongly-coupled $\nu_3$ R(3) transition.

We can also achieve resonant depopulation of the lower state of the strongly-coupled R(3) transition with a two-color experiment wherein we pump the P(3) transition at 3345.8 nm (2988.8 $cm^{-1}$). We observe a clear modulation in cavity transmission upon pumping P(3) as plotted in Fig. 5, with transient spectral features closely resembling those seen under resonant pumping of R(3) in Fig. 4c. We therefore confirm that the modulated cavity transmission reflects depletion of the $J$ = 3 ground-state population shared by both the P- and R-branch transitions.

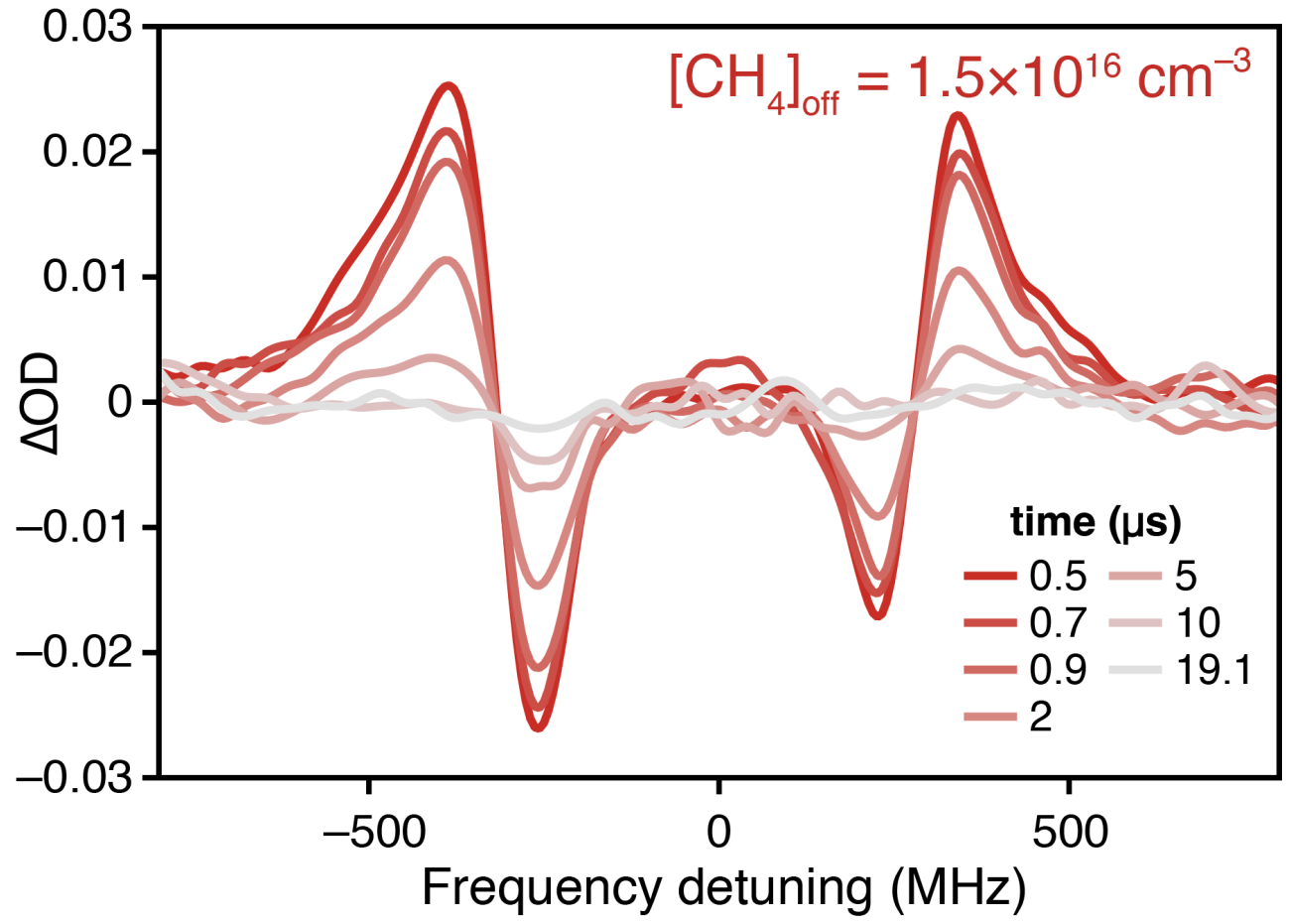

**Fig. 5.** Experimental differential cavity transmission spectra ΔOD(ν, $t_m$) at several time points $t_m$, probed in the spectral region of the R(3) transition following optical pumping of P(3). Spectra are shown with $[CH_4]_{off} = 1.5 \times 10^{16}$ cm$^{-3}$ at 295 K. The frequency detuning axis is referenced with the center of the $CH_4$ $\nu_3$ R(3) absorption band at 3057.687423 cm$^{-1}$ corresponding to 0 MHz.

We conduct several additional control experiments. We observe no transient modulation signals in empty-cavity control experiments performed without gas in the chamber, ruling out cavity-only or pump-scatter artifacts as the source of the observed ΔOD. In non-resonant control experiments, we red- and blue-detune the pump wavelength by ~5 nm from the $\nu_3$ R(3) transition at 3270.4 nm to avoid resonant pumping of any methane transitions. As expected, detuning the pump eliminates any modulation signals. We also observe no dependence of the magnitude or timescales of the transient signals on relative pump-probe polarization as detailed in Section S3.1 of the SI. We finally rule out the contributions of transient changes in cavity length to our signals as outlined in Section S3.2 of the SI.

**Modeling transient cavity transmission spectra**

To better understand the observed transient signatures, we fit our experimental data with an extension of the same classical cavity optics approach used above for linear cavity transmission spectra. We are chiefly interested in whether intracavity molecules exhibit distinct dynamics from molecules in free space, which one cannot easily recover from naive inspection of the transient cavity transmission spectra. We therefore use the analytical Fabry-Pérot cavity transmission formalism to simulate transient cavity transmission spectra and disentangle the intracavity molecular dynamics from the response of the cavity to those dynamics. Additional details for this fitting procedure are provided in Section S2.2 of the SI.

Briefly, we fit the experimental transient cavity transmission spectrum $\Delta OD(\nu, t_m)$ at each $t_m$ time point to a corresponding simulated trace

$$\Delta OD_{sim}(\nu, t_m) = -\log_{10}[\ I_{on,sim}(\nu, t_m)/I_{off,sim}(\nu, t_m)\ ] \cdot c(\nu) \quad (3)$$

where $I_{on,sim}(\nu, t_m)$ and $I_{off,sim}(\nu, t_m)$ are each calculated using eq 2. $c(\nu)$ is a Beer-Lambert factor that corrects for frequency-dependent attenuation of the probe light by extracavity $[CH_4]_{off}$ near the molecular resonance (see Section S2.2 of the SI for details). We take $L$ and $R$ as fixed inputs acquired from fitting the linear reference spectrum $I_{ref}(\nu)$ acquired before a given transient experiment, as described above. For $I_{off,sim}(\nu, t_m)$, the absorption coefficient is fixed as $\alpha_{off}(\nu) = \sigma(\nu) \cdot [CH_4]_{off}$, where $[CH_4]_{off}$ is again obtained from the fit of $I_{ref}(\nu)$. To fit $\Delta OD_{sim}(\nu, t_m)$ to experimental data, we float only the absorption coefficient used in $I_{on,sim}(\nu, t_m)$, according to $\alpha_{on}(\nu, t_m) = A(t_m) \cdot [CH_4]_{off} \cdot \sigma(\nu)$. Here, $A(t_m)$ is a scaling factor fit for each time-point $t_m$ that allows us to recover the pump-induced, time-dependent change in population in the lower state of the strongly-coupled transition, given by

$$\Delta[CH_4]_{t,\, J=3} = [1 - A(t)] \cdot [CH_4]_{off} \cdot P_{J=3} \quad (4)$$

where $P_{J=3}$ = 4.4% is the fraction of methane molecules expected to be found in the $J$ = 3 rotational level of the ground vibrational state at 295 K according to Boltzmann statistics. We fit $\Delta[CH_4]_{t,\, J=3}$ to a bi-exponential function to obtain characteristic time constants $\tau_1$ and $\tau_2$ for the population dynamics (*vide infra*).

This fitting routine yields simulated transient spectra that are in quantitative agreement with the experimental spectra acquired under both weak and strong coupling conditions, as illustrated in Fig. 6. The lefthand column of Fig. 6 shows data from a weak coupling experiment carried out with $[CH_4]_{off} = 1.4 \times 10^{15}$ cm$^{-3}$ (corresponding to the green trace in Fig. 4a), while the righthand column presents the strong coupling data with $[CH_4]_{off} = 8.0 \times 10^{15}$ cm$^{-3}$ (corresponding to the purple traces in Fig. 4). Within each column, the experimental $\Delta OD(\nu, t_m)$ spectra are precisely reproduced by $\Delta OD_{sim}(\nu, t_m)$ in both shape and magnitude for each $t_m$. We recover the dynamics of the intracavity molecules by relating the changes in $\Delta OD_{sim}(\nu, t_m)$ to changes in population in the $\nu_3$ R(3) lower state coupled transition, $\Delta[CH_4]_{t,\, J=3}$ according to eq 4. Figures 6e,f plot the extracted values of $\Delta[CH_4]_{t,\, J=3}$ in black circles as a function of time; bi-exponential fits to these data are shown in solid-colored lines. The fits are very good with $R^2>0.97$ for the two specific cases shown. Overall, quantitative agreement between the FP cavity transmission simulations and experimental data confirm that the classical cavity optics treatment accurately captures the transient spectroscopy of this system under both weak and strong coupling.

The two datasets shown in Fig. 6 are representative examples drawn from our series of time-domain measurements spanning the weak and strong coupling regimes by tuning $[CH_4]_{off}$ from 0 to $1.5 \times 10^{16}$ cm$^{-3}$. Both cases illustrated in Fig. 6 feature derivative-like Rabi contraction signatures as the population available for cavity-coupling is depleted. However, the magnitude of $\Delta OD(\nu, t_m)$ near the molecular resonance at 0 MHz detuning is greater for the weak coupling case in Figs. 6a,c, while $\Delta OD(\nu, t_m)$ remains close to 0 at 0 MHz detuning in the strong coupling case in Figs. 6b,d. Inspection of the weak coupling reference spectrum (green trace in Fig. 4a) makes it clear that the polariton peaks are not fully resolved in this case, giving the differential spectrum a singular central bleach feature. In the strong coupling case, the polariton peaks are better resolved (purple trace in Fig. 4a) and there is significant resonant attenuation of transmitted probe light at 0 MHz detuning by the high number density of intra- and extracavity molecules. As a result, the central bleach feature is visually split by R(3) absorption. In general, the observed differential signals become smaller in magnitude and noisier as $[CH_4]_{off}$ increases due to complete attenuation of probe light on resonance, complicating differential lineshapes and making individual spectral fits and extracted time dynamics less reliable for $[CH_4]_{off} > 10^{16}$ cm$^{-3}$. We report transient data for the highest methane density near $1.4 \times 10^{16}$ cm$^{-3}$ which features a maximum Rabi splitting of 637 MHz (see Fig. S4 in the SI).

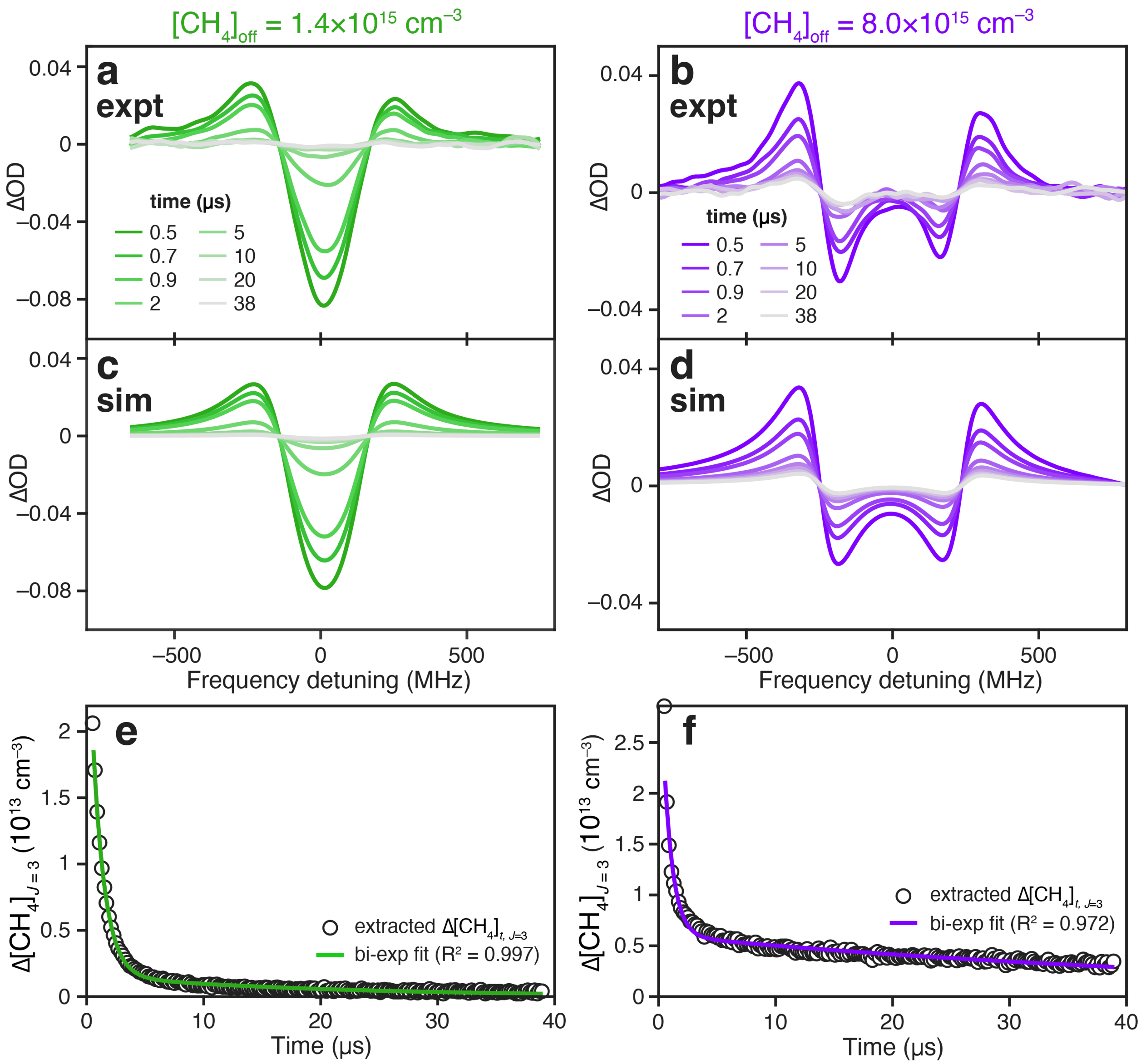

**Fig. 6. (a,b)** Experimental and **(c,d)** simulated differential cavity transmission spectra, $\Delta OD(\nu, t_m)$, at several time points $t_m$ for weak cavity-coupling with $[CH_4]_{off} = 1.4 \times 10^{15}$ cm$^{-3}$ (left column, green) and strong cavity-coupling with $[CH_4]_{off} = 8.0 \times 10^{15}$ cm$^{-3}$ (right column, purple) at room temperature (295 K). Simulated spectra result from fitting the corresponding experimental data with the differential cavity transmission expression in eq 3. The data in panel (b) is reproduced from Fig. 4c. All frequency detuning axes are referenced with the center of the $CH_4$ $\nu_3$ R(3) absorption band at 3057.687423 cm$^{-1}$ corresponding to 0 MHz. **(e, f)** The transient change in population of the lower state of the strongly coupled $\nu_3$ R(3) transition, $\Delta[CH_4]_{t,\,J=3}$, recovered from fitting of the experimental data sets in panels (a) and (b) (black circles). The time-evolution of $\Delta[CH_4]_{t,\,J=3}$ for each system is well fit with a decaying bi-exponential function, as plotted in solid-colored traces.

**Comparing transient dynamics across cavity and non-cavity experiments**

For each transient cavity transmission experiment performed with a given $[CH_4]_{off}$, we extract $\Delta[CH_4]_{t,\,J=3}$ as described above and fit its biexponential decay to obtain the fast and slow time constants, $\tau_1$ and $\tau_2$. We repeat this procedure under non-cavity conditions, with cavity mirrors removed from the chamber, to obtain control measurements of $\tau_1$ and $\tau_2$ as a function of $[CH_4]_{off}$. Representative non-cavity spectra and fits are presented in Section S3.3 of the SI. Figure 7 compares the values of $\tau_1$ and $\tau_2$ observed in cavity and non-cavity experiments as a function of $[CH_4]_{off}$. We observe qualitatively similar trends in $\tau_1$ and $\tau_2$ versus $[CH_4]_{off}$ in a data set acquired with the methane gas cooled to ~230 K, as illustrated in Fig. S5 of the SI.

We find the fast time constant $\tau_1$ to be on the order of 1 μs for the lowest methane concentrations and decrease with increasing $[CH_4]_{off}$, leveling off around ~0.5 μs (Fig. 7a). We ascribe this behavior to a combination of two processes. First, we expect the transit time of pumped molecules out of the probe volume (and unpumped molecules into the probe volume) to occur on the timescale of 0.3 μs for room temperature methane, assuming a ~200 μm cavity beam diameter and ballistic molecular motion at the low pressures used herein. We also expect rotational-rotational state changing collisions to cause transient signals to decay on the microsecond timescale.[66] Optical pumping of R(3) transiently transfers population from the $J = 3$ rotational level of the vibrational ground state to the $\nu_3$, $J = 4$ level. The bleach in R(3) will recover both as the lower $J = 3$ level is collisionally re-equilibrated with neighboring thermally-populated rotational states and as the upper $\nu_3$, $J = 4$ level is collisionally depopulated into the manifold of neighboring rovibrational states. Klaassen *et al.*[67] performed time-resolved infrared double-resonance experiments on $CH_4$ and found that the $J = 3$ sublevels of the ground vibrational state underwent rotational re-equilibration with pressure-dependent rates of 9.2 ± 2.2 $\mu s^{-1}$ $Torr^{-1}$. Several state-resolved studies of rotational dynamics in methane isotopologues report similar timescales.[68–70] We expect the lower state of R(3) to re-equilibrate on a timescale of 0.2−4 μs over the range of number densities used herein (1 to 15 × $10^{15}$ $cm^{-3}$, corresponding to 0.03 to 0.46 Torr at room temperature). It is therefore sensible that our observed $\tau_1$ is on the order of ~1 μs and decreases with increasing $[CH_4]_{off}$ due to accelerated rotational state-changing collisions at higher pressures.

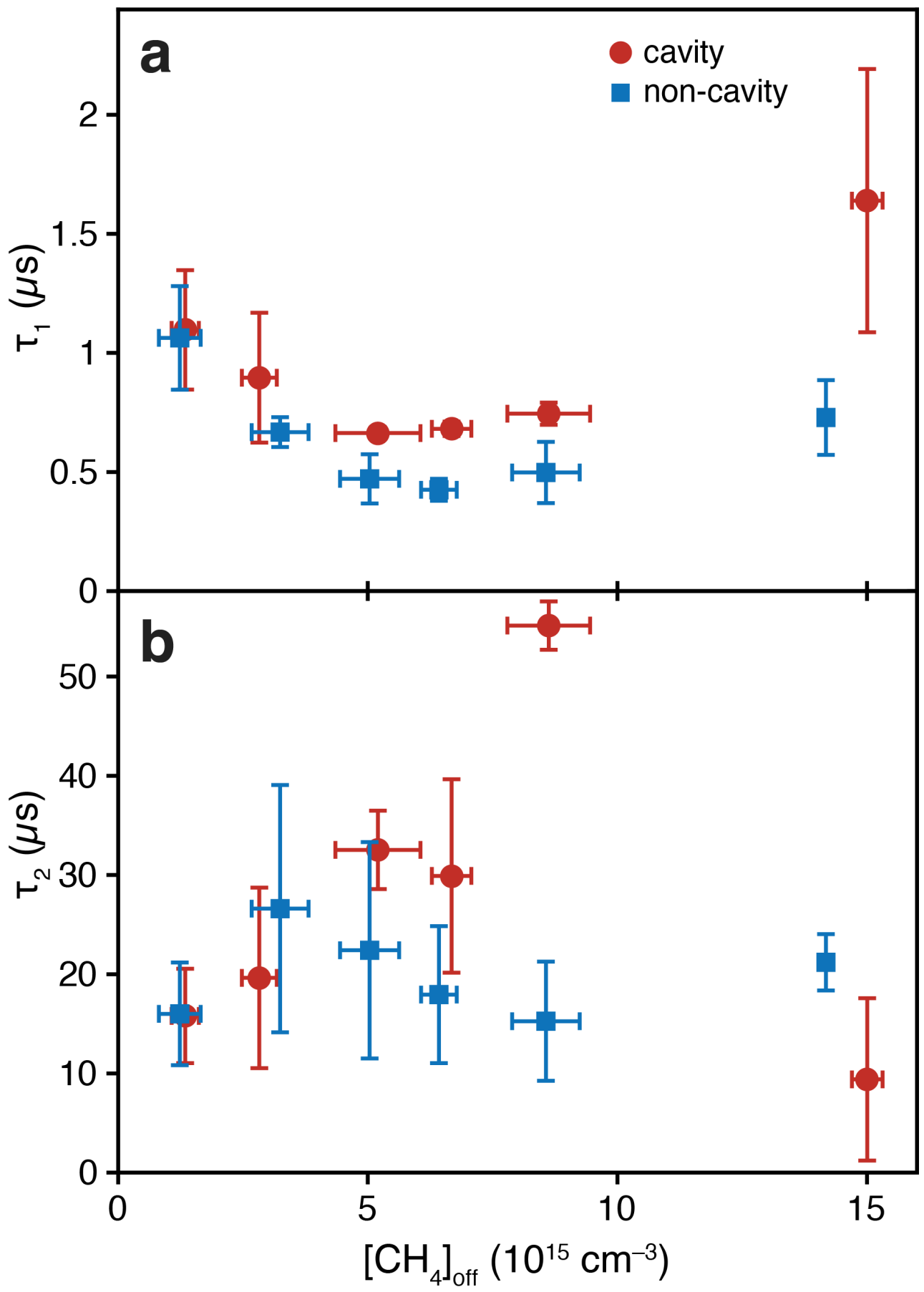

**Fig. 7.** Gas-phase recovery lifetimes for the bleach of the $\nu_3$ R(3) rovibrational transition of methane as a function of number density $[CH_4]_{off}$ at 295 K observed in cavity (red) and non-cavity (blue) experiments. **(a)** The fast time constant $\tau_1$ and **(b)** the slow time constant $\tau_2$ plotted as functions of $[CH_4]_{off}$. $\tau_1$ likely captures a combination of collisional rotational energy transfer and the transit time of molecules in and out of the probed cavity mode volume. $\tau_2$ reflects subsequent slower collisional thermalization timescales. Data points are averaged across experiments binned by similar methane number density; markers represent the bin means, while horizontal and vertical error bars denote the standard deviation ($1\sigma$) of the number densities and fitted lifetimes within each bin, respectively. The high degree of overlap between the cavity and non-cavity points across the entire density range demonstrates that the presence of the optical cavity does not perturb the underlying gas-phase molecular dynamics.

The slower time constant $\tau_2$ displays no systematic trend with molecular density (Fig. 7b), remaining on the order of 20 μs for all experiments. The substantial uncertainty in the measured $\tau_2$ values is expected given that our finite acquisition window of 20−40 μs limits our ability to measure slow decay rates. Nevertheless, the observed $\tau_2$ timescales are comparable to the 10−100 μs timescales expected for collisional vibrational-rotational and vibrational-vibrational energy transfer observed in earlier studies of methane.[71–74]

Across the full range of methane concentrations explored here, $\tau_1$ and $\tau_2$ overlap within experimental uncertainty between cavity-coupled and non-cavity conditions (see both Fig. 7 and Fig. S5). We observe no statistically significant changes in either timescale upon entering the strong coupling regime for methane densities above $[CH_4]_{off} \gtrsim 1.5 \times 10^{15}$ cm$^{-3}$. This result indicates that strong cavity coupling does not measurably perturb rovibrational collisional relaxation channels or gas transport at the coupling strengths accessed here. Our results therefore constrain any cavity-modified phenomena in this system to lie below our current experimental sensitivity or outside the range of coupling strengths accessed here.

## CONCLUSIONS AND FUTURE DIRECTIONS

The present work represents the first transient spectroscopy performed on gas-phase molecules under collective strong light-matter coupling to the best of our knowledge. We demonstrate that a classical optics treatment of our coupled cavity-molecule system can quantitatively reproduce both the linear cavity transmission spectra and the time-resolved nonlinear cavity transmission spectra, without the need to invoke a quantum optics description of polaritonic states. The intracavity population dynamics extracted from the cavity signals are statistically indistinguishable from those of non-cavity controls across the full range of coupling strengths explored here. We emphasize that even fully classical transient cavity signals often appear non-intuitive, and it is therefore essential to disentangle the molecular and cavity response before examining the dynamics of a strongly-coupled system. Only once this classical description is accounted for (and spurious optical artifacts ruled out) should one consider invoking more exotic mechanisms for cavity-modified molecular behavior.

This gas-phase strong coupling platform lends itself to several natural next steps for future work. Introducing collinear probe and pump beams along the cavity axis will enable more direct comparison to the experimental geometries used in the condensed-phase VSC dynamics literature.[21,37] The tunable pressure, density, and temperature accessible in the gas phase also make this platform well-suited to systematically explore the transition between homogeneous and inhomogeneous broadening regimes, where transient spectroscopic signatures are expected to differ substantially.[6,21,34] Probing dynamics under conditions where multiple molecular transitions are cavity-coupled[26] would allow us to search for evidence of cavity-mediated vibrational energy transfer, which has been reported in condensed-phase systems under VSC.[55]

Finally, we have recently established electronic strong coupling of the rovibronic transitions of gas-phase molecular iodine ($I_2$).[35] Applying our transient pump-probe methodology to $I_2$, perhaps with the addition of time-resolved fluorescence readout, would offer a promising route to probe quantum-state-specific excited-state dynamics under electronic strong coupling.

This work underscores the importance of accounting for classical optical cavity signals when examining molecular dynamics under vibrational strong coupling. As the field continues to investigate whether and how strong coupling modifies chemical dynamics, the quantitative spectral clarity accessible in gas-phase systems offers a clean testbed for isolating true polaritonic contributions from conventional molecular photophysics.

## ASSOCIATED CONTENT

### Data Availability Statement

Experimental details, additional plots, and experimental data are available from the authors upon reasonable request.

### Supporting Information

Experimental methods, simulation and data processing details, and additional experimental cavity transmission spectra and simulations (PDF).

## AUTHOR INFORMATION

### Corresponding Author

* Marissa L. Weichman – Department of Chemistry, Princeton University, Princeton, New Jersey 08544, United States; orcid.org/0000-0002-2551-9146; email: weichman@princeton.edu

### Authors

Jane C. Nelson – Department of Chemistry, Princeton University, Princeton, New Jersey 08544, United States; https://orcid.org/0000-0002-2560-2775

Trevor H. Wright – Department of Chemistry, Princeton University, Princeton, New Jersey 08544, United States; https://orcid.org/0000-0002-9330-8631

Madeline Rohde – Department of Chemistry, Princeton University, Princeton, New Jersey 08544, United States; https://orcid.org/0009-0004-0192-5862

Neo Lin – Department of Chemistry, Princeton University, Princeton, New Jersey 08544, United States; https://orcid.org/0000-0002-0535-564X

**Funding**

This research was supported by the National Science Foundation CAREER award CHE-2238865 and startup funds provided by Princeton University. J.C.N. acknowledges support from a National Defense Science and Engineering Graduate research fellowship under Award No. FA9550-23-F-0014. M.R. acknowledges support from the Princeton University Department of Chemistry and the Peter N. Curtin, Class of 2008 Memorial Fund.

The authors declare no competing financial interest.

## ACKNOWLEDGMENTS

The authors acknowledge the use of the Micro/Nano Fabrication Center (MNFC), a core shared-use facility of the Princeton Materials Institute (PMI), for cavity mirror fabrication.

# Supporting Information:
# Rovibrational dynamics in strongly-coupled gas-phase methane

*Jane C. Nelson, Trevor H. Wright, Madeline Rohde, Neo Lin, and Marissa L. Weichman**

Department of Chemistry, Princeton University, Princeton, New Jersey 08544, United States

* weichman@princeton.edu

**Contents**

## S1. Additional experimental methods

### S1.1. Optical cavity design and vacuum chamber

Our vacuum chamber and optical set-up are shown in Fig. 2a of the main text. We evacuate the chamber with a turbo pump (Pfeiffer) backed by a roughing scroll pump (Edwards). The chamber reaches a base pressure of $6.6 \times 10^{-4}$ torr when evacuated at room temperature and $3.3 \times 10^{-5}$ torr when the system is liquid nitrogen cooled. The flow of room temperature methane ($CH_4$) gas (Airgas) into the chamber is set with a mass flow controller (Alicat Scientific) and can be varied between 0 and 1000 sccm. $CH_4$ gas is delivered toward the center of the chamber through a copper gas line. Anti-reflection coated $CaF_2$ windows (Edmund) provide optical access to the chamber along both pump and probe axes.

Inside the vacuum chamber, we construct a two-mirror Fabry-Pérot optical cavity by mounting mirrors in standard optical mirror mounts (Thorlabs Polaris) that are clamped to the floor of the chamber ~1 cm apart. One of the mirrors is glued to a ring-shaped piezoelectric chip (pzt) for cavity length stabilization as described below in Section S1.3. We fabricate our mirrors in the Princeton Micro/Nanofabrication Center cleanroom using an electron beam evaporator (Angstrom Engineering Nexdep) to deposit gold on the concave faces of plano-concave ½" diameter $CaF_2$ substrates. We use −1.05±0.02 cm radius of curvature (*ROC*) substrates (EKSMA Optics) for the near-confocal cavity. We coat the substrates with a nominal 20±2 nm layer of gold to achieve the desired mirror reflectivity. We target a reflectivity of 95% to yield cavity linewidths comparable to the expected molecular transition linewidths.

To enable cooling for a subset of experiments, the input gas line is wound around a screw mounted to a ~1.5 L capacity liquid nitrogen ($LN_2$) dewar that is itself mounted in the vacuum chamber. The dewar and gas line temperatures are monitored with K-type thermocouples (Omega). Both thermocouples read 100 K before gas flow, and up to 110 K during gas flow. We report a much higher ~230 K translational temperature of the methane based on the observed Doppler linewidths, due to the abundance of room temperature surfaces in the chamber and the absence of the buffer gas cell used in our previous experiments.[1,2]

### S1.2. Cavity transmission spectroscopy and frequency calibration

Our laser table layout is shown in Fig. 2b of the main text. We use a continuous-wave (cw) distributed feedback interband cascade laser (ICL, Nanoplus) to measure the mid-infrared methane and optical cavity transmission spectra. The ICL operates near the 3270.4 nm (3057.7 $cm^{-1}$) target transitions of $CH_4$ with a sub-10 MHz instantaneous linewidth and several milliwatts of output power. The ICL central wavelength can be tuned over 3262 to 3278 nm (3050 to 3065 $cm^{-1}$) using a current driver (Thorlabs LDC201CU) and a temperature controller (Thorlabs TED200C). The ICL is protected from optical feedback by an optical isolator (Faraday Photonics). To set or sweep the ICL frequency, we use a function generator (BK Precision) to apply a DC offset or triangle wave to the modulation port of the current driver. The ICL light is split into three beams using an ~88% reflective homemade Au flat and a 50:50 mid-infrared beamsplitter (Thorlabs BSW511). The highest power beam (beam 1 in Fig. 2b) is aligned through the vacuum chamber to measure the cavity transmission, while beams 2 and 3 are used for absolute and relative frequency calibration. For the cavity transmission beam path, we use a pair of anti-reflection coated plano-convex $CaF_2$ lenses to mode-match the beam to the $TEM_{00}$ cavity mode. We control the polarization of beam 1 before it enters the cavity with a mid-IR half-wave plate (Thorlabs WPLH05M-3500). We measure the intensity of light transmitted through the cavity with a $LN_2$-cooled HgCdTe detector with a 53 ns time resolution specified by the manufacturer (Kolmar). We measure non-cavity transmission spectra of methane in the chamber by removing the cavity mirrors.

To obtain a linear cavity transmission spectrum, we sweep the ICL frequency over 30 GHz using a 1.1 Vpp, 10 Hz triangle wave fed to the modulation port of the current driver. For absolute calibration of the frequency axis, a portion of the 3270 nm beam (beam 2 in Fig. 2b) is directed through a homebuilt room-temperature methane vapor cell. For relative frequency calibration, the beam (beam 3 in Fig. 2b) is directed through a germanium (Ge) etalon (Light Machinery, nominal length 50.8 mm). Both the vapor cell and Ge etalon transmission spectra are recorded with InAsSb detectors (Thorlabs). The cavity, vapor cell, and Ge etalon transmission spectra are collected simultaneously to calibrate the frequency axis of the cavity transmission. All ICL beams are incident on detectors read out with the same oscilloscope (Tektronix MDO3024). Data processing and calibration procedures are described further below in Section S2.1.

### S1.3. Cavity length stabilization and tunability

To reduce vibrational noise and drift over time, a side-of-line lock to a fixed reference laser is used to stabilize the cavity length. Figure 2b depicts this locking scheme. A piezoelectric chip (Thorlabs PA44M3KW) glued to one of the cavity mirrors enables remote modulation of the cavity length. We use a metrology-grade tunable 1550 nm cw external cavity diode laser (ECDL, RIO ORION, 4 kHz linewidth) as a frequency reference. The ECDL is coupled coaxially onto beam 1 of the ICL to traverse the cavity, using Ge windows (Thorlabs WG91050-C9) that act as longpass dichroic mirrors to transmit the 3270 nm ICL beam while reflecting the 1550 nm ECDL light. We detect the intensity of the ECDL transmitted through the cavity with an amplified Ge photodetector (Thorlabs PDA30B2). The ECDL side-of-line error signal is processed with a proportional-integral (PI) loop filter (New Focus LB1005), then fed back to the cavity piezo with an open loop piezo driver (Thorlabs MDT694B). The cavity length can thus be stabilized to ensure the side of a cavity fringe is resonant with the reference laser. The cavity length can also be systematically stepped by applying a voltage offset to the output of the piezo driver.

### S1.4. Pulsed nanosecond pump laser

We use narrowband mid-IR pulsed light as the pump for time-resolved experiments. To generate this light, we begin with a seeded, pulsed Nd:YAG laser (Amplitude DLS8010) with a 10 Hz repetition rate, a pulse duration of >6 ns, and a spectral bandwidth fwhm of 0.003 $cm^{-1}$ (0.09 GHz). We use both the 1064.5 nm Nd:YAG fundamental and its 532.2 nm second harmonic generated in a deuterated potassium dihydrogen phosphate (KD*P) crystal. The Nd:YAG second harmonic is used to pump a dye laser (Sirah Lasertechnik Cobra-Stretch) loaded with 4-dicyanomethylene-2-methyl-6-p-dimethylaminostyryl-4H-pyran (DCM) laser dye dissolved in ethanol to produce light at 635.7 nm. The spectral linewidth of the dye laser output is specified as <0.066 $cm^{-1}$ (2.0 GHz) fwhm, set by a 2400 lines/mm grating in the oscillator cavity. We send both the dye laser output and the Nd:YAG fundamental to an optical parametric amplifier near-infrared system (OPANIR, Sirah Lasertechnik) to generate pulsed mid-IR light. First, the 635.7 nm and 1064.5 nm beams are overlapped in a lithium niobate ($LiNbO_3$) crystal to produce 1578.2 nm light in a difference frequency mixing (DFM) process. This 1578.2 nm light is then combined with additional 1064.5 nm light in a second $LiNbO_3$ crystal to produce the desired 3270.4 nm pulses via optical parametric amplification (OPA). For a subset of experiments, we pump the $\nu_3$ P(3) transition of methane at 3345.8 nm by tuning the dye laser wavelength to ~633 nm. The nonlinear processes described above yield pulsed mid-IR light with a linewidth of ~0.07 $cm^{-1}$ (2.0 GHz). The frequency of the mid-IR light is tunable via simultaneous computer control of the dye laser oscillator grating, the DFM crystal angle, and the OPA crystal angle. We record the wavelengths of the Nd:YAG fundamental and dye laser output before each experiment to within 0.001 nm with a wavemeter (Bristol Instruments) to confirm that the resulting mid-IR light is resonant with the methane transition of interest.

The nonlinear OPANIR processes yield 10 mJ/pulse of mid-IR light, albeit with a fairly uneven beam profile. We spatially filter this beam with a telescope and an iris as it is routed towards the chamber. Approximately 3 mJ/pulse is ultimately available at the chamber as pump light. We align the pump beam through the chamber orthogonal to the cavity axis and check that the beam does not clip on any surfaces using a mid-IR beam profiler (WinCamD-IR-BB). The beam passes through a side port of the chamber fitted with a recessed anti-reflection coated $CaF_2$ window, and is focused vertically with a cylindrical lens to maximize overlap with the cavity mode volume. The pump beam is chopped with a phase-locked mechanical chopper (Thorlabs MC2000B and MC1F2) for shot-to-shot pump-on/pump-off data collection.

We use a digital delay generator (DDG, Stanford Research Systems DG645) to synchronize the experiment and collect time-resolved data. The DDG triggers the Nd:YAG laser flash lamp and Q-switch, sets the chopper phase, and directs oscilloscope collection from the three cw ICL detectors for cavity transmission and frequency calibration. Custom LabVIEW software is used as the main platform for experiment control, including to adjust the pump and probe laser frequencies, record chamber pressure and temperature, and monitor the status of the cavity lock. LabVIEW is also used to acquire and save data in hdf5 files for downstream processing in MATLAB. We digitize and route signals with a National Instruments PCIe-6351 multifunction I/O device connected through a BNC-2120 shielded terminal block. Data processing methods are further described in Section S2 below.

## S2. Modeling and data processing

### S2.1. Frequency-domain data processing

We calibrate the frequency axis of each reference spectrum $I_{\mathrm{ref}}(\nu)$ using the simultaneously-acquired vapor cell and Ge etalon transmission spectra (beams 2 and 3 in Fig. 2b and described above in Section S1.2). We subtract the linear baseline of the etalon spectrum, extract the positions of extrema, and use the known etalon length to make an initial guess at its *FSR*. We anchor this relative frequency axis to the absolute frequency of the $CH_4$ vapor cell transmission features. We iterate on the effective etalon *FSR* to minimize the error of the $CH_4$ peak positions until they are within 3 MHz of their known positions. We then apply this calibrated frequency axis to our cavity transmission spectra.

As introduced in eq 2 of the main text, we model cavity transmission spectra using the classical optics expression for the fractional intensity of light transmitted by a Fabry-Pérot cavity composed of two identical mirrors and filled with a homogeneous dielectric medium.[3,4] Constructive interference produces peaks in the cavity transmission spectrum whenever $\delta(\nu)$, the round-trip phase accrued by light traveling in the cavity, is equal to an integer multiple of $2\pi$, which minimizes the denominator of eq 2. When $n(\nu) = n_0$ is constant, transmission maxima are evenly spaced by the cavity free spectral range, $FSR = c/2n_0L$. However, near an absorption feature, $n(\nu)$ takes on a dispersive lineshape and can yield additional constructive interference conditions, including the polaritonic cavity mode splitting that occurs under strong coupling.

Before fitting, we scale each experimental reference cavity transmission spectrum $I_{\mathrm{ref}}(\nu)$ such that the intensity of cavity modes far from methane absorption lines are normalized to 1. Because the cavity only fills a portion of the probe pathlength in the chamber we find we must also account for absorption of light by extracavity methane. For all simulations reported herein, we model the transmission through both cavity and chamber by scaling the cavity transmission intensity $I_T(\nu)/I_0$ from eq 2 by Beer's law according to:

$$\frac{I_{\mathrm{chamber}}(\nu)}{I_0} = \frac{I_T(\nu)}{I_0} \times \exp[-\sigma(\nu) \cdot [\mathrm{CH_4}] \cdot \ell] \qquad \text{(S1)}$$

where $\ell = 14$ cm is the extracavity pathlength in the chamber. We fit $I_{\mathrm{ref}}(\nu)$ with eq S1 to extract $L$, $R$, and $[CH_4]$. Note that we do not constrain $T$ in this fit, as $T$ impacts the intensity of cavity transmission features but not their lineshapes; we discard any information about absolute cavity transmission intensity in normalizing $I_{\mathrm{ref}}(\nu)$. We generate initial guesses for each fit as follows: $L$ from the experimental *FSR*, taking $L = c/2n_0 \cdot FSR$ with $n_0 = 1$; $[CH_4]$ from the known gas flow rate; and $R = 0.94$ from mirror characterization. The intracavity absorption coefficient $\alpha(\nu)$ is calculated as $\alpha(\nu) = \sigma(\nu) \cdot [CH_4]$, where the absorption cross section $\sigma(\nu)$ is obtained using the HITRAN linelist for methane[5] processed in PGOPHER[6] to account for pressure and temperature broadening. $n(\nu)$ is then derived from $\alpha(\nu)$ via the Kramers-Kronig relation.

Simulated fits of various representative experimental linear cavity transmission spectra $I_{ref}(\nu)$ are shown below in Fig. S1. By fitting experimental data with eq S1, we obtain accurate cavity parameters and initial methane number densities, which serve as input for fitting time-resolved data, as described below in Section S2.2. The fitted cavity length $L$ is less precisely constrained than $R$ and [$CH_4$] since small imperfections in the frequency calibration of $I_{ref}(\nu)$ propagate directly into $L$. We do not interpret fit-to-fit variation in $L$ as evidence of real changes in cavity length across a set of experiments, since the cavity lock is maintained continuously and does not usually break or reset between measurements at different $CH_4$ flow rates.

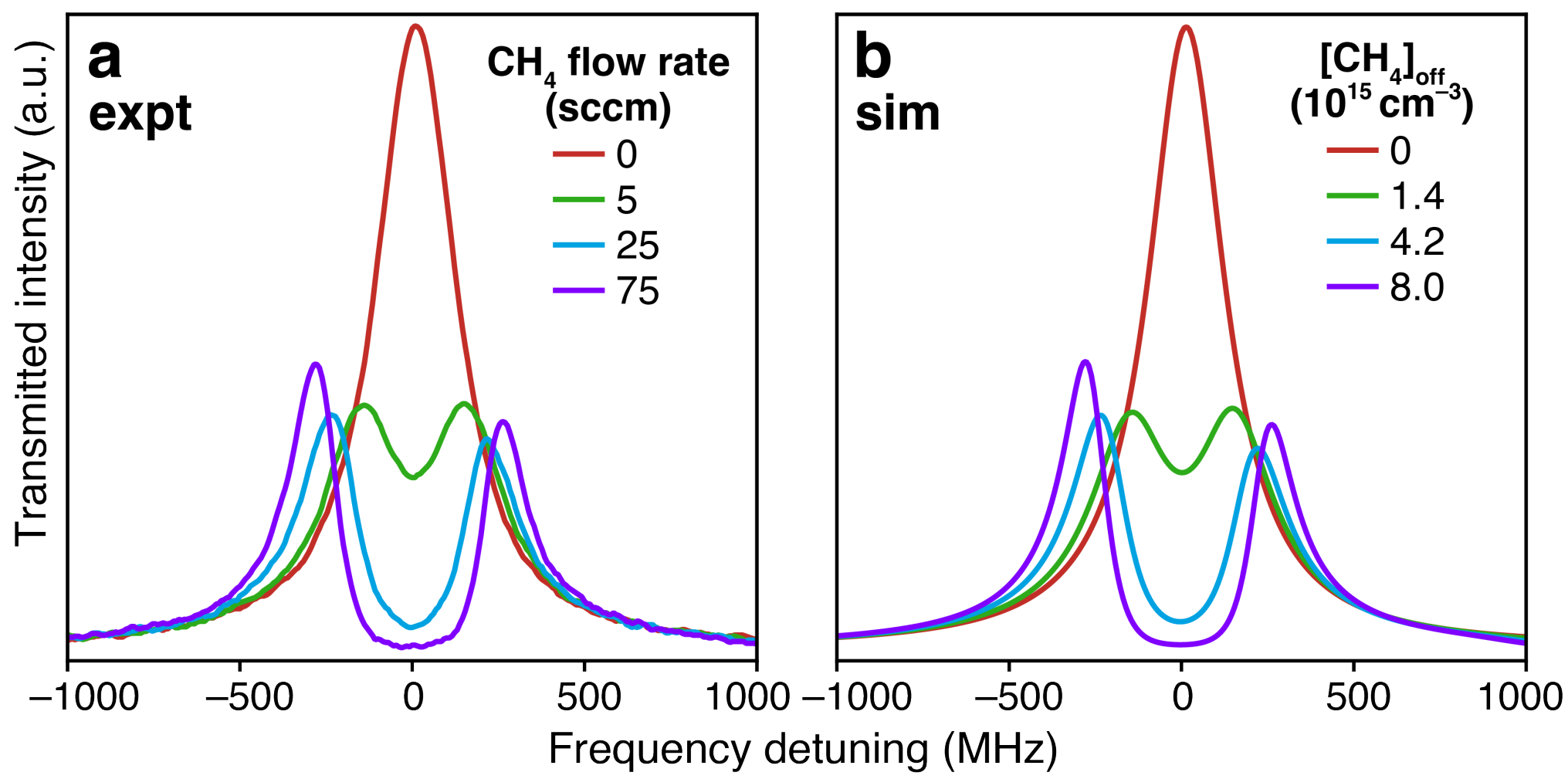

**Fig. S1.** Representative fits of linear cavity transmission spectra, $I_{ref}(\nu)$. **(a)** Experimental linear cavity transmission spectra for $CH_4$ flow rates of 0, 5, 25, and 75 sccm, reproduced from Fig. 4a. **(b)** Simulated spectra that result from fitting each experimental spectrum in panel (a) to eq S1, yielding methane number densities of $[CH_4]_{off}$ = 0, 1.4, 4.2, and $8.0 \times 10^{15}$ cm$^{-3}$, respectively. All data shown here are acquired at room temperature. The frequency detuning axes are referenced with the center of the $CH_4$ $\nu_3$ R(3) absorption band at 3057.687423 cm$^{-1}$ corresponding to 0 MHz.

### S2.2. Time-domain data processing

We generate time-resolved differential cavity transmission traces $\Delta OD(\nu_n, t) = -\log_{10}[\, I_{on}(\nu_n, t) / I_{off}(\nu_n, t)\,]$ (eq 1 of the main text) from the pump-on and pump-off transmission traces acquired at each cw ICL frequency setpoint $\nu_n$. The ICL frequency is stepped in ~33 MHz steps across a ~2 GHz spectral window near the $\nu_3$ R(3) methane line. We acquire 50−75 pairs of pump-on/pump-off time traces at each $\nu_n$ setpoint. Each time trace is collected with an oscilloscope (Tektronix) as 10,000 points over a 20 or 40 µs acquisition window (2 or 4 ns/point, respectively), finer than the ~53 ns time resolution of the HgCdTe cavity transmission detector. We calibrate the frequency axis of the $\Delta OD(\nu, t)$ differential spectra by examining a temporal slice through the $I_{off}(\nu_n, t)$ data with $t < 0$, and mapping the observed spectral features onto those of the calibrated $I_{ref}(\nu)$ reference spectrum.

Traces collected at each ICL setpoint are averaged together, then smoothed with a Savitzky-Golay method (3$^{rd}$ order, 25-point frame) to generate time-dependent cavity transmission traces with pump on [$I_{on}(\nu_n, t)$] and pump off [$I_{off}(\nu_n, t)$]. Before smoothing the time traces, we establish the pump-probe overlap time $t$=0 when $d[I_{on}(\nu_n, t)]/dt$ exceeds a noise threshold. This noise threshold is taken to be five times the standard deviation of the first 50 time-points of $d[I_{on}(\nu_n, t)]/dt$, well before the pump arrives. Note that in this processing, time traces with too much frequency noise or complete saturation of the probe light are discarded from the $I_{on}(\nu_n, t)$ and $I_{off}(\nu_n, t)$ data sets. We discard individual traces when the ICL frequency is unstable, identified by deviation from the average baseline transmission for $t < 0$ for that setpoint frequency. This instability usually results from the ICL control electronics picking up electrical noise from the Nd:YAG flashlamps. We discard the data from an entire ICL setpoint when less than a third of the pump-on or pump-off traces are retained and there are not enough traces to reliably average and process. This situation most often occurs for runs performed at high methane concentrations ($[CH_4] > 10^{16}$ molecules $cm^{-3}$) when all ICL probe light is absorbed near the molecular resonance.

Because our time traces are sampled much more densely than the cavity transmission photodetector's time resolution, we further smooth the time-domain data by averaging together the differential time traces $\Delta OD(\nu_n, t)$ acquired for each $\nu_n$ setpoint in discrete bins of 5 neighboring time points. We hereafter label these discrete time bins as $t_m$. Scattered pump light can obscure the desired cavity transmission signals for the first few hundred nanoseconds of each time trace so we begin defining these bins 500 ns after $t = 0$ for cavity experiments.

We fit the time-resolved cavity transmission spectrum $\Delta OD(\nu, t_m)$ at each $t_m$ time bin using the classical Fabry-Pérot cavity model described above in Section S2.1. We construct each simulated differential frequency-domain spectrum with eq 3 in the main text. We find the term $c(\nu) = \exp[-\sigma(\nu) \cdot [CH_4]_{off} \cdot d]$ in eq 3 necessary to account for extracavity attenuation of the probe light. $d = 10$ cm provides the best fit across all $[CH_4]_{off}$ conditions. This $c(\nu)$ correction term is somewhat unintuitive as the extracavity attenuation in the pump-on and pump-off conditions is expected to cancel out per eq S1. We believe the extreme attenuation of probe light near the molecular resonance necessitates the correction, which enables quantitative agreement between simulation and experiment in terms of both the lineshapes and magnitudes of transient signals.

In simulating the $I_{off,sim}(\nu, t_m)$ pump-off spectra using eq 2 of the main text, we take the intracavity absorption coefficient to be $\alpha(\nu) = \sigma(\nu) \cdot [CH_4]_{off}$. The resulting $I_{off,sim}(\nu, t_m)$ spectra therefore, by construction, match the reference spectrum $I_{ref}(\nu)$. To fit the pump-induced modulation in pump-on traces, we float the scaling of the intracavity absorption coefficient $\alpha_{on}(\nu, t_m) = A(t_m) \cdot [CH_4]_{off} \cdot \sigma(\nu)$. We seed the fit with an initial guess for $A(t_m)$ using a biexponential decay with literature time constants.[7] We then refine $A(t_m)$ by minimizing the difference between $\Delta OD_{sim}(\nu, t_m)$ and the experimental $\Delta OD(\nu, t_m)$ at each $t_m$ time point.

## S3. Control experiments

### S3.1. Polarization dependence

All data presented in the main text is collected with pump and probe lasers that feature parallel vertical polarization. Here, we consider tuning the ICL probe laser polarization relative to that of the pump in order to examine whether polarization has any impact on our measurements of methane population relaxation dynamics (Fig. S2). We use a half-wave plate to rotate the probe polarization from 0˚ (vertical) to 90˚ (horizontal) with respect to the pump polarization. Representative differential transmission spectra for cavity experiments with $[CH_4]_{off} = 1.5 \times 10^{16}$ $cm^{-3}$ at select time delays are presented in Fig. S2a-c. These spectra illustrate that the overall spectral shape and amplitude of transient signatures are preserved as a function of relative pump-probe polarization angle. Correspondingly, we observe no systematic change or trend in the short time constant, $\tau_1$, extracted from fitting data acquired across pump-probe polarization angles under both cavity and non-cavity conditions (Fig. S2d-f). This polarization independence is fully consistent with the spherical-top symmetry of methane as well as the rapid rotational reorientation of the molecules that is too fast for our time resolution.

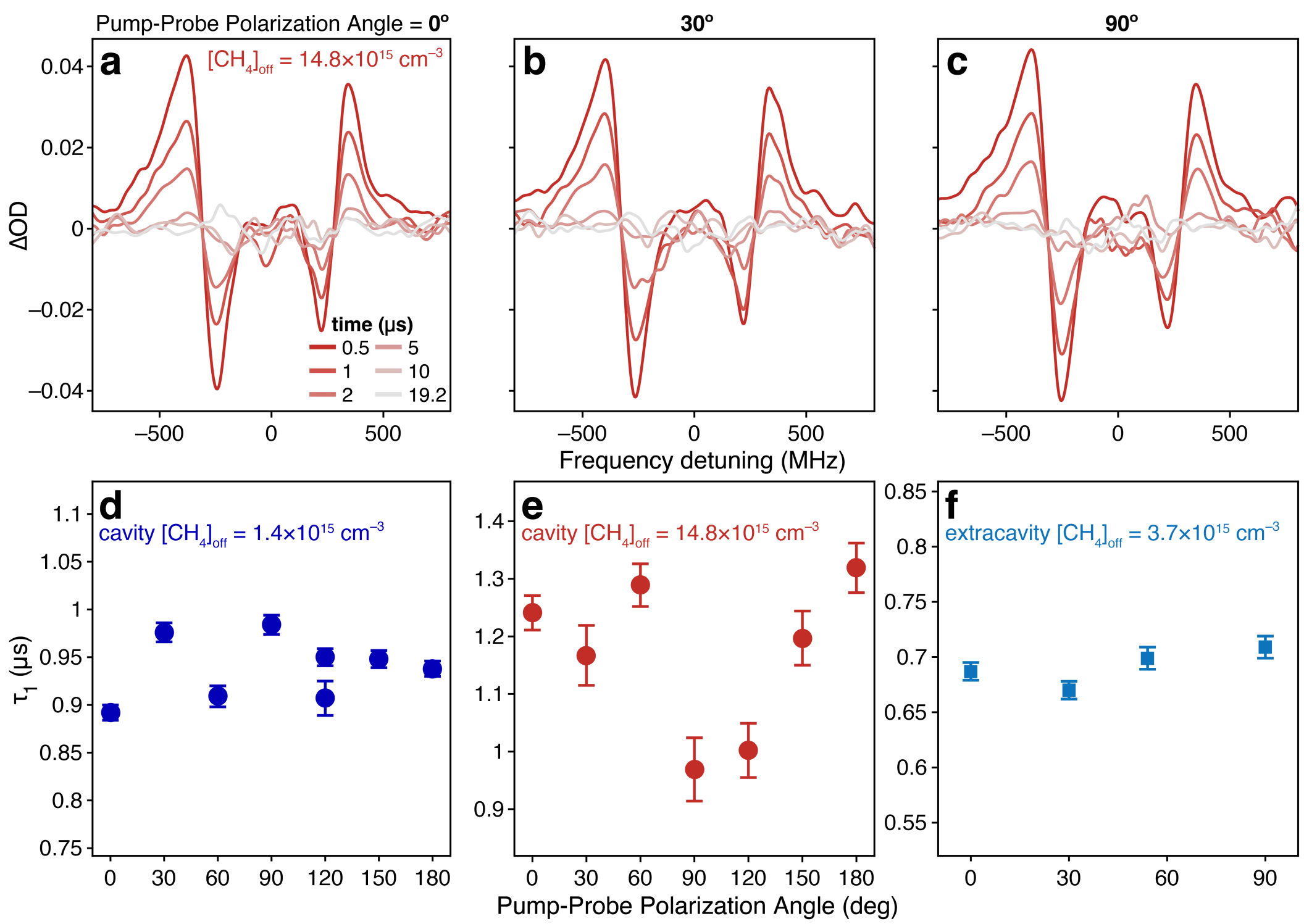

**Fig. S2.** Polarization dependence of methane population relaxation. **(a-c)** Differential cavity transmission spectra ($\Delta OD(\nu, t_m)$) at select pump-probe delay times for a cavity with methane at a 200 sccm flow rate ($[CH_4]_{off} = 1.5 \times 10^{16}$ $cm^{-3}$), at pump-probe polarization angles of **(a)** 0°, **(b)** 30°, and **(c)** 90°. **(d-f)** Extracted short time constants ($\tau_1$) for experiments in a cavity with methane at a **(d)** 5 sccm ($[CH_4]_{off} = 1.4 \times 10^{15}$ $cm^{-3}$) and **(e)** 200 sccm ($[CH_4]_{off} = 1.5 \times 10^{16}$ $cm^{-3}$) flow rate, and **(f)** without a cavity with methane at a 20 sccm ($[CH_4]_{off} = 3.7 \times 10^{15}$ $cm^{-3}$) flow rate at different probe laser polarization angles, with respect to the pump laser vertical polarization. Vertical error bars in (d-f) denote the 95% confidence interval of the bi-exponential fit to $\Delta[CH_4]_{t, J=3}$. Probe laser polarization is changed via turning a half-wave plate prior to the cavity.

### S3.2. Cavity length stability during pumping

To confirm that the observed pump-induced $\Delta OD(\nu, t_m)$ modulation does not arise from a transient change in cavity length, we performed a control fit in which $L$ was allowed to float independently between the pump-on and pump-off components of $\Delta OD_{sim}(\nu, t_m)$, defined in Section S2.2 above. This fitting procedure did not yield an improved fit. We also note that transient changes in cavity length serve to shift the frequencies of the cavity modes rather than modify their shape or linewidth. Therefore, a transient change in cavity length would not be expected to produce the semi-symmetric, derivative-like $\Delta OD(\nu, t_m)$ lineshapes we observe near the coupled transition for all $[CH_4]_{off}$ studied herein.

### S3.3. Non-cavity spectroscopy

We conduct non-cavity control experiments to benchmark the $CH_4$ lineshape and transient response to pumping in free space. As in the cavity experiments, the orthogonal pump intersects ~1 cm of the probe laser path length. The probe beam shape is unchanged from cavity experiments; the cavity mirrors are simply removed from the beam path. First, we collect linear transmission reference spectra of methane in the chamber (Fig. S3a). We fit these spectra with the Beer-Lambert law to extract $[CH_4]_{off}$, again using PGOPHER absorption cross-section data specific to the Doppler and pressure broadening conditions of each flow rate. Data acquisition, averaging, smoothing, and $t = 0$ determination follow the same procedure described for cavity experiments in Section S2.2. Representative transient time traces for various ICL probe frequencies $\Delta OD(\nu_n, t)$ are shown in Fig. S3b, while differential transmission spectra $\Delta OD(\nu, t_m)$ at various pump-probe delay times are shown in Fig. S3c for $[CH_4] = 0$ to $8 \times 10^{15}$ cm$^{-3}$.

We rely on simple Beer-Lambert attenuation to extract state-specific population dynamics in non-cavity experiments. As in Section S2.2, we float a time-dependent scaling factor $A(t_m)$ at each pump-probe delay to reproduce the observed $\Delta OD(\nu, t_m)$. We exclude ICL setpoints from analysis when the probe laser transmission is too strongly attenuated to reliably extract a differential signal. We track the population dynamics of the lower state of the transition of interest, $\Delta[CH_4]_{t, J=3} = [1 - A(t)] \cdot [CH_4]_{off} \cdot P_{J=3}$ (eq 4 of the main text), and fit $\Delta[CH_4]_{t, J=3}$ to a bi-exponential decay to extract $\tau_1$ and $\tau_2$, as shown in Fig. S3d. This procedure is repeated for a series of $CH_4$ flow rates to generate the non-cavity $\tau_1$ and $\tau_2$ values reported as a function of $[CH_4]_{off}$ in Fig. 7 of the main text and in Fig. S5 below.

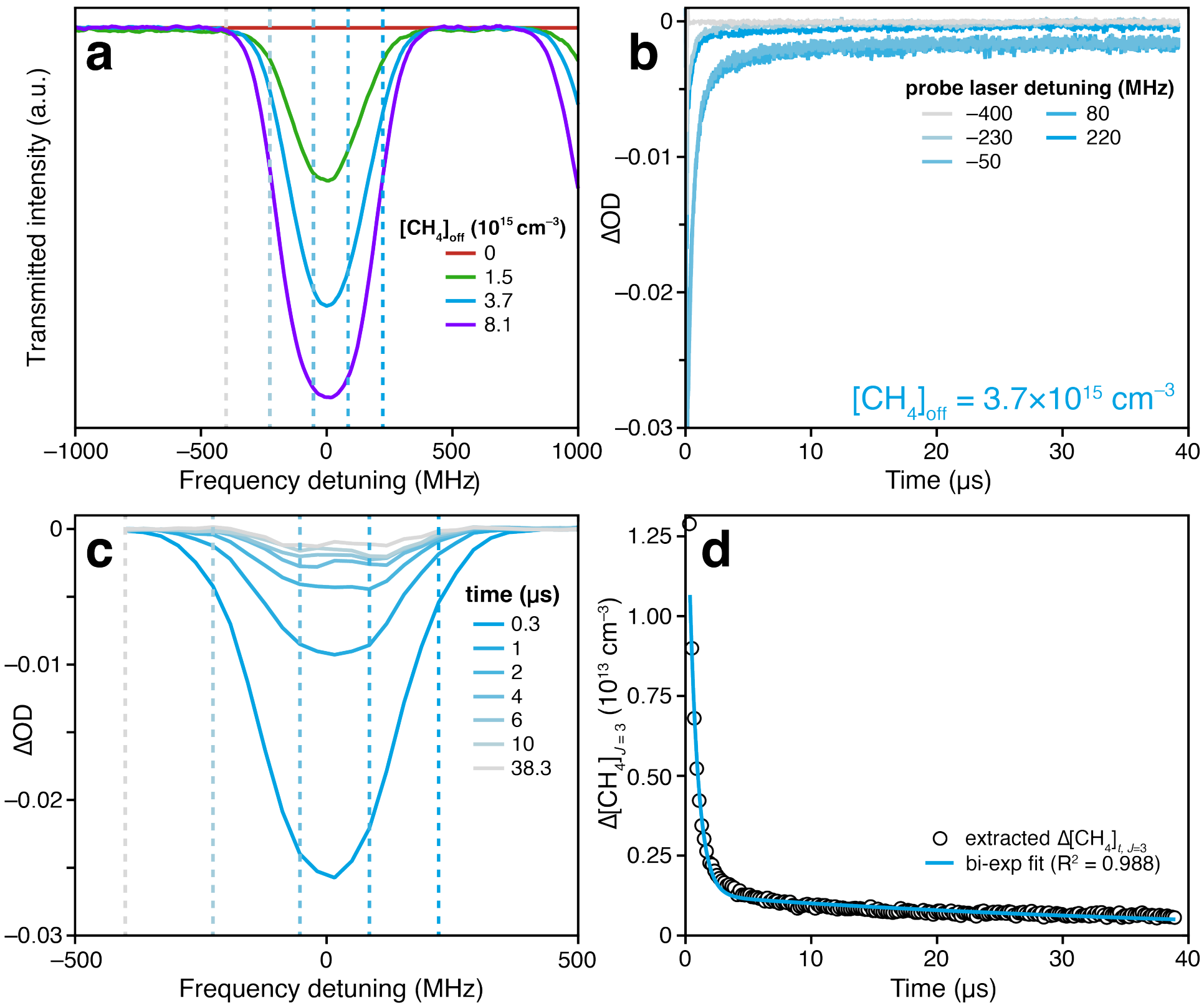

**Fig. S3.** Non-cavity control measurements. **(a)** Transmission spectra of the cavity-free chamber under systematic tuning of the initial methane number density $[CH_4]_{off}$ via increasing flow rate from 0, 5, 20, to 75 sccm. **(b)** Differential optical density $\Delta OD(v_n, t)$ time traces at probe frequencies $v_n$ near −400, −230, −50, 80, and 220 MHz for $[CH_4]_{off} = 3.7 \times 10^{15}$ cm$^{-3}$ corresponding to the blue linear spectrum in panel (a). **(c)** $\Delta OD(v, t_m)$ spectra at several pump-probe time delays $t_m$. **(d)** Resulting state-specific population depletion $\Delta[CH_4]_{t,\, J=3}$ with bi-exponential fits. Frequency axes are referenced with the $CH_4$ $\nu_3$ R(3) absorption band at 3057.687423 cm$^{-1}$ corresponding to 0 MHz.

## S4. Additional figures

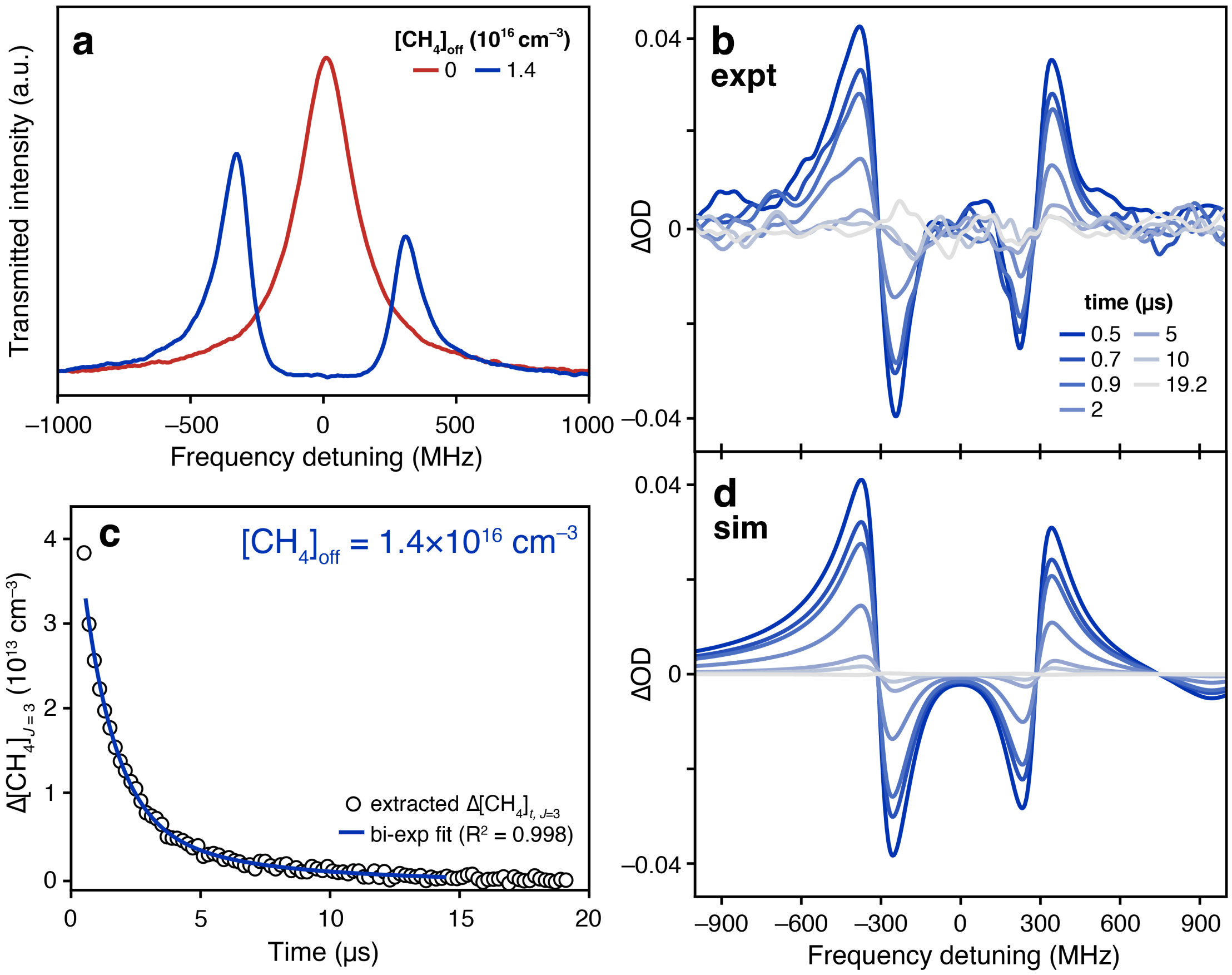

**Fig. S4.** Linear and transient cavity transmission spectra for the higher flow rate of $[CH_4]_{off} = 1.4 \times 10^{16}$ cm$^{-3}$ at room temperature. **(a)** Linear cavity transmission spectra comparing the empty cavity condition and the $[CH_4]_{off} = 1.4 \times 10^{16}$ cm$^{-3}$ strongly-coupled condition, which features a larger Rabi splitting than the data in Figs. 4 and 6 of the main text. **(b)** Experimental and **(d)** simulated differential optical density spectra $\Delta OD(\nu, t_m)$ at time points $t_m$ following optical pumping of R(3). Fits here are less reliable than for the conditions shown in the main text due to greater attenuation of both pump and probe light by extracavity gas. **(c)** The resulting population depletion $\Delta[CH_4]_{t,\,J=3}$ (black circles) and bi-exponential fit (solid blue line). Frequency detuning axes in panels (a), (b), and (d) are referenced with the center of the $CH_4$ $\nu_3$ R(3) absorption band at 3057.687423 cm$^{-1}$ corresponding to 0 MHz.

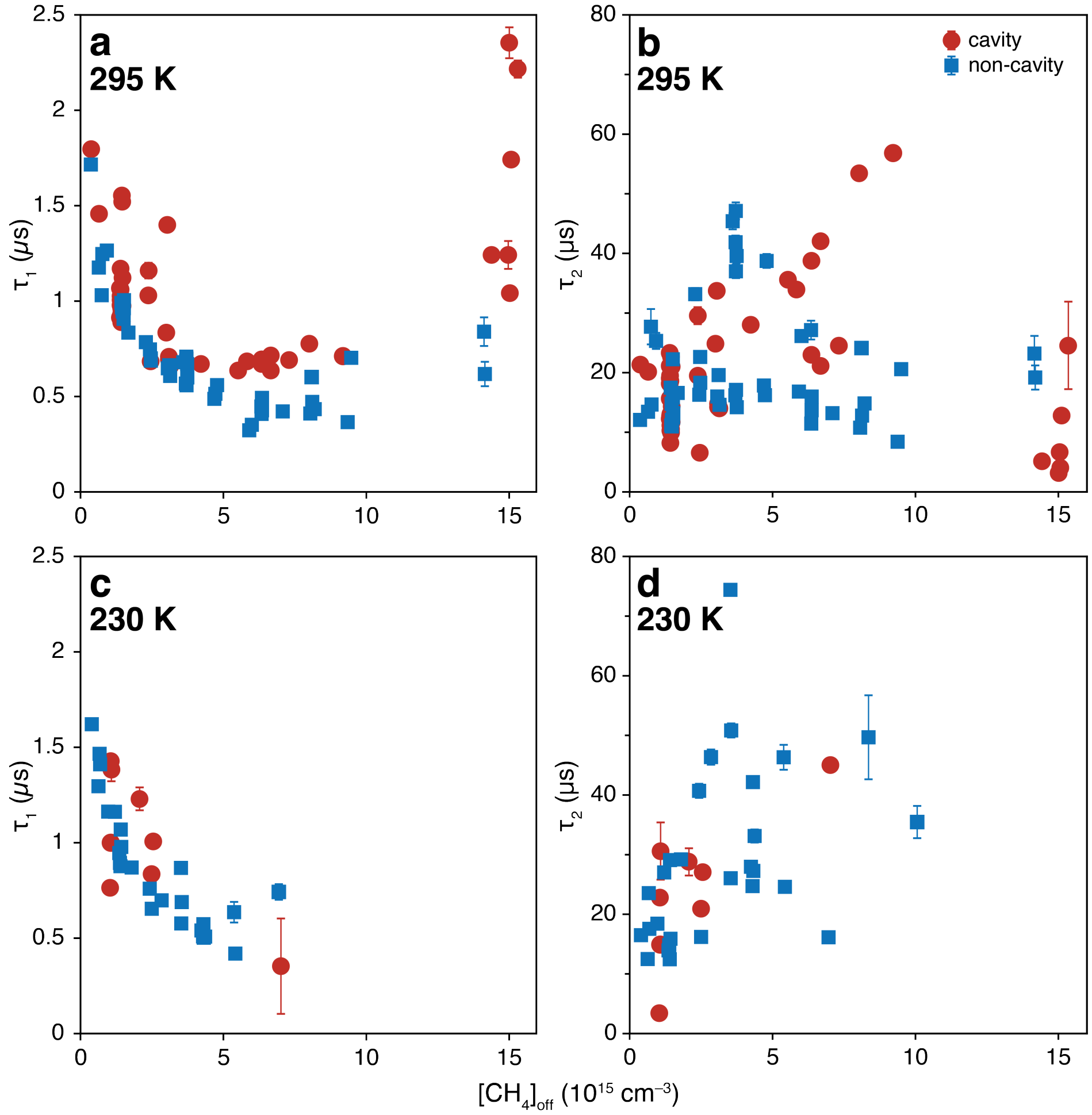

**Fig. S5**. Gas-phase recovery lifetimes for the bleach of the $\nu_3$ R(3) rovibrational transition of methane as a function of initial number density $[CH_4]_{off}$ near 295 and 230 K under cavity (red) and non-cavity (blue) configurations at room temperature (a,b; 295 K) and reduced temperature (c,d; ~230 K). **(a,c)** The fast relaxation lifetime component, $\tau_1$, extracted from bi-exponential fits of the state-specific population depletion over time. **(b,d)** The slow lifetime component, $\tau_2$. Each data point represents the values extracted from a single transient experiment performed at a fixed flow rate. The vertical error bars denote the 95% confidence interval of the bi-exponential fit to $\Delta[CH_4]_{t,\,J=3}$; horizontal error bars are omitted but $[CH_4]_{off}$ uncertainties are typically ~5% of the plotted value. The overlap of the intracavity and non-cavity trends across the entire density range demonstrates that the presence of the optical cavity does not perturb the underlying molecular kinetics or gas transport. The inconsistencies at high number densities arise from difficulties in fitting transient dynamics when the pump and probe lasers are highly attenuated by extracavity gas.